\documentclass[12pt]{article} 

\usepackage[a4paper,total={18cm,27cm}]{geometry}

\usepackage[utf8]{inputenc}
\usepackage{amsmath}
\usepackage{amsfonts}

\usepackage{graphicx}
\usepackage{array}
\usepackage{caption}

\usepackage{xcolor}
\colorlet{shadecolor}{yellow!20}

\usepackage{authblk}

\usepackage{hyperref}

\usepackage{comment}   

\newcommand{\NI}{\vspace{0.2cm}\noindent}

\begin{document}


\title{Convergent Evolution in Neural Representation Space: Emergent Order in Deep Belief Networks}


\author[1,2,3,4]{Patrick Krauss}
\author[1,2,3,4]{Achim Schilling}
\author[1]{Andreas Maier}
\author[2]{Thomas Kinfe}
\author[1]{Claus Metzner}

\affil[1]{\small Cognitive Computational Neuroscience Group, Pattern Recognition Lab, Friedrich-Alexander-University Erlangen-Nürnberg (FAU), Germany}
\affil[2]{\small Neuromodulation and Neuroprosthetics, University Hospital Mannheim, University Heidelberg, Germany}
\affil[3]{\small Neuroscience Lab, University Hospital Erlangen, Germany}
\affil[4]{\small BG Clinic Ludwigshafen, Germany}

\maketitle


\begin{abstract}
\NI Deep Belief Networks (DBNs) learn hierarchical generative models without class supervision. Here, we ask whether this purely unsupervised process nevertheless organizes internal representations according to the unknown data classes. We analyze successive layers of DBNs trained on MNIST, Fashion-MNIST, and KMNIST using the Generalized Discrimination Value (GDV), supervised probes applied only after training, a reconstruction-based measure of abstraction distance, effective dimensionality, and free sample generation. Remarkably, class-specific clustering generally increases with depth across datasets and network widths, although no label information is available during DBN training. Control experiments show that this effect depends on the learned feature structure and cannot be explained by random transformations, weight marginals, dimensionality reduction, or sigmoid saturation. The first hidden layers also frequently make class identity more accessible to linear and nonlinear probes. With greater depth, representations become increasingly compact and prototype-like as neurons acquire correlated feature directions. At the same time, GDV and probe accuracy reveal complementary aspects of class structure: improved average clustering can coexist with reduced accessibility for a few difficult class pairs. These findings demonstrate that layer-wise generative learning can spontaneously uncover and progressively amplify class-related structure in unlabeled data.
\end{abstract}

\newpage

\section{Introduction}

\NI The modern resurgence of deep learning is closely connected to the introduction of efficient training procedures for Deep Belief Networks (DBNs). Hinton, Osindero, and Teh showed that a deep generative model could be constructed greedily, one Restricted Boltzmann Machine (RBM) at a time, and subsequently fine-tuned for a supervised task \cite{hinton2006fast}. This layer-wise pretraining strategy provided a practical route into parameter regimes that were difficult to reach by direct optimization of deep networks and helped establish unsupervised representation learning as a central idea in the emerging field \cite{bengio2007greedy}. The broader foundational importance of energy-based neural networks and learning algorithms associated with John Hopfield and Geoffrey Hinton was recognized by the 2024 Nobel Prize in Physics \cite{nobel2024physics}. DBNs therefore occupy an important historical position: they connected probabilistic generative modeling, statistical mechanics, and the trainability of deep neural architectures at a formative stage of modern machine learning.

\NI Despite this historical role, DBNs subsequently receded from the center of deep-learning research. Improved parameter initialization and activation functions alleviated vanishing-gradient problems \cite{glorot2010understanding}, while residual connections enabled very deep discriminative networks to be trained directly by backpropagation \cite{he2016deep}. At the same time, increasing data availability, specialized hardware, and end-to-end optimization shifted attention toward architectures tailored to particular application domains. Transformer networks now provide the dominant framework for many sequence and language tasks \cite{vaswani2017attention}, and diffusion models have become a powerful approach to high-quality sample generation \cite{ho2020denoising}. In this landscape, greedy layer-wise pretraining is usually no longer required merely to make a deep classifier trainable. As a consequence, DBNs are often treated primarily as a historical stepping stone, and comparatively little attention is paid to the internal representational changes produced by their unsupervised training itself.

\NI Yet several properties make DBNs unusually useful model systems for questions at the interface of machine learning, computational neuroscience, and statistical physics. First, their learning is unsupervised: no class labels or task-specific output errors are required. Second, the network is trained locally and layer by layer, allowing the contribution of each newly learned transformation to be examined separately. Third, every RBM defines an explicit probability distribution and energy function, linking neuronal activations to a tractable energy-based description \cite{ackley1985learning,hinton2002training}. Fourth, the learned hierarchy can be used both as an encoder and as a generative model. These features do not make DBNs literal models of biological neural circuits: symmetric weights, equilibrium sampling, and the separation into positive and negative learning phases remain substantial abstractions. Nevertheless, local correlation-based updates, stochastic neuronal states, recurrent associative dynamics, and learning without externally supplied labels make them attractive conceptual systems for investigating how structured internal representations can emerge from environmental statistics alone.

\NI Our group has pursued a related program of analyzing information processing inside neural networks without restricting the analysis to final task performance. We introduced the Generalized Discrimination Value (GDV) as a non-invasive measure of how strongly labeled data classes separate within an arbitrary representation, permitting comparisons across layers and architectures without modifying the analyzed network \cite{schilling2021quantifying}. In recurrent and Boltzmann-type networks, we further studied spontaneous information flux and identified efficiently measurable correlational signatures of state-to-state information transmission \cite{metzner2024quantifying}. We investigated how noise can release recurrent networks from restricted attractor subsets through recurrence resonance \cite{metzner2024recurrence}, how neuronal nonlinearity and dynamical regimes determine whether reservoir representations become linearly decodable \cite{metzner2025nonlinear}, and how biologically motivated structural regularities such as Dale-type output polarity, reciprocity, and modularity affect recurrent dynamics and computation \cite{metzner2025organizational}. Across these studies, a recurring theme is that useful information processing cannot be inferred from architecture or output accuracy alone: it requires direct measurement of representational geometry, dimensionality, accessibility, and dynamics.

\NI Here, we return to DBNs from this representational perspective and ask what their greedy, completely unsupervised learning procedure does to data as they pass through the hierarchy. In particular, we distinguish three properties that are often implicitly conflated: preservation of a specific input, separation and extractability of class-related information, and the ability to generate plausible samples in a free-running mode. We quantify layer-wise class separation using the GDV and assess class accessibility with linear, finite Gaussian-kernel, and RBF support-vector-machine probes. In parallel, we measure deterministic abstraction distance, operationalized by the conventional reconstruction mean squared error, and the effective dimensionality of each representation, and decompose dimensionality loss into variance heterogeneity and correlation-induced neuronal redundancy. Free generation and data-grounded reconstruction provide complementary visual tests of the learned generative models. All labels are withheld during DBN training and are introduced only after training for evaluation.

\NI We perform these analyses across MNIST, Fashion-MNIST, and KMNIST, multiple hidden-layer widths, independently initialized networks, and a set of controls targeting initialization, contrastive-divergence depth, weight structure, nonlinearity, stochastic sampling, and label assignment. This design allows us to test whether progressively improved GDV reflects a robust unsupervised reorganization rather than a peculiarity of one data set or training run. Our results reveal a depth-dependent trade-off: early layers discard pixel-level detail while making class information more accessible, whereas deeper layers increasingly compress the representation into correlated, partially redundant feature directions and eventually reduce supervised readout accuracy. Free generative behavior follows a distinct trajectory, demonstrating that abstraction distance, discriminative accessibility, and generative quality are related but non-equivalent properties of an unsupervised deep model.

\section{Methods}

\subsection{Datasets}

\NI Experiments were performed on three established image-classification benchmarks with identical technical formats but different visual statistics. Each dataset contains 60,000 training images and 10,000 test images, represented as $28\times28$ grayscale arrays and assigned to ten classes. MNIST comprises handwritten Arabic digits from 0 to 9 \cite{lecun1998gradient}. Fashion-MNIST was designed as a drop-in replacement for MNIST and contains images of ten clothing categories, including garments, footwear, and bags \cite{xiao2017fashion}. Kuzushiji-MNIST (KMNIST) contains ten classes of cursively written Japanese Kuzushiji characters and provides a visually more complex handwritten-symbol task while retaining the MNIST data format \cite{clanuwat2018deep}.

\NI We used the official training and test partitions throughout. Pixel intensities were converted to floating-point values in $[0,1]$ and images were flattened to 784-dimensional input vectors; no data augmentation was applied. DBNs were trained exclusively on the image vectors. Dataset labels were withheld from all RBM and DBN training procedures and were used only after training to construct class-balanced GDV subsets and to fit or evaluate the supervised probe models described below. Unless specified otherwise, quantitative test-set measures were evaluated on the complete official test partition.

\subsection{Generalized Discrimination Value (GDV)}

\NI The Generalized Discrimination Value (GDV) was used to quantify the separability of labeled data classes in the representations generated by successive DBN layers \cite{schilling2021quantifying}. The GDV maps a set of labeled points to a single scalar by comparing the typical distances between points belonging to the same class with those between points belonging to different classes. By construction, it is invariant under global translations, component-wise linear rescaling, and permutations of the representation coordinates (i.e., neuron indices). Its normalization and class-wise averaging further make values comparable across representations with different dimensionalities, different numbers of samples per class, and different numbers of classes, properties referred to as dimensionality, class-size, and class-number invariance, respectively. Class labels were used exclusively for this post-hoc evaluation and were not available to the DBN during training.

\NI Consider $N$ representation vectors $\mathbf{x}_n=(x_{n,1},\ldots,x_{n,D})$ in a $D$-dimensional space, with labels assigning each point to one of $L$ classes $C_l$. To make the measure invariant under translation and component-wise scaling, each dimension was standardized according to
\begin{equation}
s_{n,d}=\frac{1}{2}\frac{x_{n,d}-\mu_d}{\sigma_d},
\end{equation}
where $\mu_d$ and $\sigma_d$ denote the mean and standard deviation of dimension $d$ over all evaluated samples. Dimensions with zero variance were set to zero after standardization.

\NI For a class $C_l$ containing $N_l$ points, the mean intra-class distance was defined as
\begin{equation}
\bar d(C_l)=\frac{2}{N_l(N_l-1)}
\sum_{i=1}^{N_l-1}\sum_{j=i+1}^{N_l}
\left\lVert\mathbf{s}^{(l)}_i-\mathbf{s}^{(l)}_j\right\rVert_2 .
\end{equation}
Likewise, the mean inter-class distance between classes $C_l$ and $C_m$ was
\begin{equation}
\bar d(C_l,C_m)=\frac{1}{N_lN_m}
\sum_{i=1}^{N_l}\sum_{j=1}^{N_m}
\left\lVert\mathbf{s}^{(l)}_i-\mathbf{s}^{(m)}_j\right\rVert_2 .
\end{equation}
The GDV was then calculated as
\begin{equation}
\mathrm{GDV}=\frac{1}{\sqrt{D}}
\left[
\frac{1}{L}\sum_{l=1}^{L}\bar d(C_l)
-\frac{2}{L(L-1)}\sum_{l=1}^{L-1}\sum_{m=l+1}^{L}\bar d(C_l,C_m)
\right].
\end{equation}
The factor $1/\sqrt{D}$ reduces the dependence on representation dimensionality. Values near zero indicate strongly overlapping classes, whereas increasingly negative values indicate progressively stronger class separation. For all comparisons within a data set, the GDV was computed on the same fixed, class-balanced evaluation subset comprising 200 samples per class. Thus, variation across trained networks reflects differences in their learned representations rather than differences in the samples used for GDV estimation.

\subsection{Abstraction distance and prototypical shift}

\NI To quantify how strongly a reconstruction departs from a specific input at a given network depth, we measured its mean squared difference from the original image. Mathematically, this is the conventional mean squared reconstruction error. Conceptually, however, we refer to this quantity as the
\emph{abstraction distance} in the present study: an increasing value can reflect not only loss of recoverable information, but also the removal of idiosyncratic detail as the reconstruction moves toward a more class-typical form. We use the term \emph{prototypical shift} for this underlying transformation toward prototype-like structure. Abstraction distance denotes the magnitude of the input--reconstruction difference, whereas prototypical shift denotes its directed interpretation; the latter therefore requires visual or class-conditional evidence and cannot be inferred from the scalar distance alone. For an input vector $\mathbf{x}_n\in[0,1]^{D_0}$ and a DBN truncated after hidden layer $l$, the deterministic forward representation was computed by successively applying the hidden-unit activation probabilities of the first $l$ RBMs. This representation was then propagated back to the visible layer in reverse order using the corresponding visible-unit activation probabilities, yielding the reconstruction $\hat{\mathbf{x}}_n^{(l)}$. No stochastic binary states were sampled during this evaluation.

\NI The abstraction distance at depth $l$, operationalized as reconstruction MSE, was defined in the original input space as
\begin{equation}
D_{\mathrm{abs}}^{(l)}=
\frac{1}{N D_0}\sum_{n=1}^{N}\sum_{d=1}^{D_0}
\left(x_{n,d}-\hat{x}_{n,d}^{(l)}\right)^2 .
\end{equation}
It was evaluated over the complete test set. A smaller value indicates that more input-level detail can be recovered through the first $l$ layers, whereas a larger value indicates a stronger transformation away from the individual input. This transformation is interpreted as a prototypical shift only where the reconstructed images demonstrably become more class-typical; otherwise, the same increase may reflect indiscriminate information loss or distortion. For the untransformed input representation ($l=0$), reconstruction is the identity operation and the abstraction distance was therefore defined as zero. This quantity is an evaluation measure and is distinct from the layer-local contrastive-divergence objective used to train the individual RBMs.

\subsection{Effective dimensionality}

\NI The effective dimensionality of each layer representation was quantified by the participation ratio of its covariance spectrum. Let $\mathbf{h}_n^{(l)}\in\mathbb{R}^{D_l}$ denote the deterministic activation-probability vector produced by layer $l$ for test sample $n$, and let
\begin{equation}
\mathbf{C}^{(l)}=\frac{1}{N-1}\sum_{n=1}^{N}
\left(\mathbf{h}_n^{(l)}-\bar{\mathbf{h}}^{(l)}\right)
\left(\mathbf{h}_n^{(l)}-\bar{\mathbf{h}}^{(l)}\right)^{\!\top}
\end{equation}
be the sample covariance matrix. From its non-negative eigenvalues $\lambda_1,\ldots,\lambda_{D_l}$, the effective dimensionality was calculated as
\begin{equation}
D_{\mathrm{eff}}^{(l)}=
\frac{\left(\sum_{i=1}^{D_l}\lambda_i\right)^2}
{\sum_{i=1}^{D_l}\lambda_i^2} .
\end{equation}
This measure estimates how many mutually orthogonal variance directions contribute substantially to the representation. It approaches one when nearly all variance is concentrated along a single direction and approaches the algebraic rank when variance is distributed uniformly across all active directions. If the representation has zero total variance, $D_{\mathrm{eff}}$ was defined as zero. The measure is invariant under a common rescaling of all activations and was evaluated on the complete test set without using class labels.

\subsection{Decomposition of effective dimensionality}

\NI To distinguish a loss of dimensionality caused by unequal single-neuron variances from a loss caused by statistical dependencies between neurons, we derived several complementary quantities from the same layer-wise covariance matrix $\mathbf{C}^{(l)}$. All quantities were evaluated on the deterministic activation-probability representations of the complete test set. For compactness, the layer superscript is omitted below. Let $D$ be the number of units in the representation, $C_{ij}$ the entries of $\mathbf{C}$, and $\lambda_1,\ldots,\lambda_D$ its eigenvalues. We define the total covariance energy as
\begin{equation}
Q=\operatorname{tr}(\mathbf{C}^2)
=\sum_{i=1}^{D}\lambda_i^2
=\sum_{i=1}^{D}\sum_{j=1}^{D}C_{ij}^2 .
\end{equation}

\NI First, a \emph{variance-only effective dimension} was calculated after discarding all off-diagonal covariances while retaining the observed variance of every unit:
\begin{equation}
D_{\mathrm{var}}
=\frac{\left(\sum_{i=1}^{D}C_{ii}\right)^2}
{\sum_{i=1}^{D}C_{ii}^2} .
\end{equation}
This is the participation ratio of the diagonal matrix $\operatorname{diag}(\mathbf{C})$. It becomes small when the total activity variance is concentrated in only a few neurons, but it is unaffected by correlations among different neurons. Comparing $D_{\mathrm{var}}$ with the actual $D_{\mathrm{eff}}$ therefore separates variance heterogeneity from covariance-induced redundancy.

\NI The corresponding \emph{off-diagonal covariance fraction} was defined as
\begin{equation}
R_{\mathrm{cov}}
=\frac{\sum_{i\neq j}C_{ij}^2}{\sum_{i,j}C_{ij}^2}
=1-\frac{\sum_i C_{ii}^2}{Q}
=1-\frac{D_{\mathrm{eff}}}{D_{\mathrm{var}}} .
\end{equation}
It measures the fraction of squared covariance-matrix energy contributed by dependencies between distinct neurons. A value of zero indicates a diagonal covariance matrix, whereas values approaching one indicate that off-diagonal covariances dominate the second-order structure. The final identity follows because $D_{\mathrm{eff}}=(\operatorname{tr}\mathbf{C})^2/Q$ and $D_{\mathrm{var}}=(\operatorname{tr}\mathbf{C})^2/\sum_i C_{ii}^2$.

\NI As a more directly interpretable measure of pairwise dependence, Pearson correlations were computed for all pairs of non-constant units. For the set $\mathcal{A}=\{i:C_{ii}>10^{-12}\}$ of $M=|\mathcal{A}|$ variable units,
\begin{equation}
r_{ij}=\frac{C_{ij}}{\sqrt{C_{ii}C_{jj}}},
\qquad
\overline{|r|}=\frac{2}{M(M-1)}
\sum_{\substack{i<j\\i,j\in\mathcal{A}}}|r_{ij}| .
\end{equation}
The absolute value treats positive and negative coupling equally because either reduces statistical independence. Unlike $R_{\mathrm{cov}}$, which weights covariance contributions by their magnitude, $\overline{|r|}$ assigns equal weight to every valid neuron pair.

\NI To quantify the concentration of population activity along its single dominant direction, we measured the fraction of variance explained by the first principal component,
\begin{equation}
F_{\mathrm{PC1}}=\frac{\lambda_{\max}}{\sum_i\lambda_i} .
\end{equation}
An increase in $F_{\mathrm{PC1}}$ indicates that a growing share of the representation varies along one common activity mode and thus provides a spectral complement to the participation ratio.

\NI Finally, two scale-dependent activity measures were included. The \emph{total activity variance} was the trace of the covariance matrix,
\begin{equation}
V_{\mathrm{tot}}=\operatorname{tr}(\mathbf{C})
=\sum_i C_{ii}=\sum_i\lambda_i ,
\end{equation}
which quantifies the overall spread of the representation across test examples. The \emph{mean single-neuron activity standard deviation} was
\begin{equation}
\overline{\sigma}_{\mathrm{unit}}
=\frac{1}{D}\sum_{i=1}^{D}\sqrt{C_{ii}} .
\end{equation}
These two quantities test whether dimensionality collapse is accompanied by a general contraction of the population representation or by progressively weaker variation of individual units. In contrast to $D_{\mathrm{eff}}$, $D_{\mathrm{var}}$, $R_{\mathrm{cov}}$, $\overline{|r|}$, and $F_{\mathrm{PC1}}$, they are not invariant under a common rescaling of all activations. For any quantity whose denominator vanished, the value was defined as zero.

\subsection{Weight-space redundancy}

\NI To test whether functional redundancy among neuronal activities originated from redundancy among the learned features, we compared the incoming weight vectors of neurons within each RBM layer. Let $\mathbf{w}_i^{(l)}$ denote the incoming weight vector of neuron $i$ in layer $l$. Pairwise directional similarity was quantified by the cosine
\begin{equation}
s_{ij}^{(l)}=
\frac{\mathbf{w}_i^{(l)\top}\mathbf{w}_j^{(l)}}
{\|\mathbf{w}_i^{(l)}\|_2\,\|\mathbf{w}_j^{(l)}\|_2} .
\end{equation}
We report both the mean absolute cosine over all neuron pairs and the mean, over neurons, of the largest absolute cosine to any other neuron in the same layer. The first quantity measures global alignment of the weight population, whereas the second is sensitive to near-duplicate feature partners even when they constitute only a small fraction of all possible pairs. Absolute values treat parallel and antiparallel feature directions as directionally redundant.

\NI The dimensionality of the complete weight population was calculated after normalizing every weight vector to unit length. Let $\widetilde{\mathbf{W}}^{(l)}$ be the matrix containing these normalized vectors as columns and let $\sigma_k$ be its singular values. The normalized weight-matrix dimension was defined as
\begin{equation}
D_W^{(l)}=
\frac{\left(\sum_k \sigma_k^2\right)^2}
{\sum_k \sigma_k^4} .
\end{equation}
This is the participation ratio of the eigenvalue spectrum of $\widetilde{\mathbf{W}}^{(l)\top}\widetilde{\mathbf{W}}^{(l)}$ and estimates the number of substantially occupied weight directions independently of differences in neuronal weight norms.

\NI Finally, we related structural and functional similarity across all neuron pairs within each layer. Signed weight cosines $s_{ij}^{(l)}$ were compared with the corresponding signed Pearson activity correlations $r_{ij}^{(l)}$ using a Pearson correlation across pairs. In parallel, the absolute values $|s_{ij}^{(l)}|$ and $|r_{ij}^{(l)}|$ were compared using a Spearman rank correlation. The signed statistic tests whether parallel and antiparallel weight directions predict activity coupling of the corresponding sign, whereas the absolute statistic tests whether greater directional similarity generally predicts stronger functional dependence. Activity correlations were evaluated over all 10,000 MNIST test images.

\subsection{Layer-wise classification probes}

\NI To quantify how readily class information could be extracted from each frozen representation, three supervised probe models were trained independently for the input and every hidden layer. Class labels were used only for fitting and evaluating the probes and were never supplied to the DBN or used to modify its representations. For every probe, feature standardization was fitted exclusively on the corresponding training representation and then applied unchanged to the test representation. Unless stated otherwise, probes used all 60,000 MNIST training examples and were evaluated on the complete, disjoint set of 10,000 test examples. No test result was used for hyperparameter selection. Probe accuracies were averaged across the same five independently initialized DBNs used in the main analyses.

\paragraph{Linear logistic probe.}
\NI The linear probe consisted of a multinomial logistic classifier acting directly on the standardized layer representation $\mathbf{z}^{(l)}$,
\begin{equation}
p(y=c\mid\mathbf{z}^{(l)})=
\frac{\exp\!\left(\mathbf{w}_c^{\top}\mathbf{z}^{(l)}+b_c\right)}
{\sum_{k=0}^{9}\exp\!\left(\mathbf{w}_k^{\top}\mathbf{z}^{(l)}+b_k\right)} .
\end{equation}
The weights were optimized by stochastic gradient descent with an $L_2$ penalty, regularization parameter $\alpha=10^{-4}$, a maximum of 100 epochs, tolerance $10^{-3}$, shuffling, and averaged SGD coefficients. No layer-specific hyperparameter tuning was performed. This probe tests whether class information is accessible through linear decision boundaries in the frozen representation.

\paragraph{Gaussian-kernel probe.}
\NI The Gaussian probe provided a controlled nonlinear readout while keeping the nonlinear feature map independent of class labels. After training-set standardization, 200 centers $\mathbf{u}_j^{(l)}$ were fitted separately in every layer by MiniBatchKMeans without labels (batch size 2,048, maximum 100 iterations, and three initializations). Each standardized representation was then mapped to radial features
\begin{equation}
\phi_j^{(l)}(\mathbf{z})=
\exp\!\left[-\frac{\|\mathbf{z}-\mathbf{u}_j^{(l)}\|_2^2}
{2\left(\sigma^{(l)}\right)^2}\right],
\qquad j=1,\ldots,200 .
\end{equation}
One shared bandwidth was used within each layer. Its squared value was defined as the median squared distance from a fixed random sample of up to 10,000 training representations to their nearest center,
\begin{equation}
\left(\sigma^{(l)}\right)^2=
\operatorname{median}_{n}\min_j
\|\mathbf{z}_n^{(l)}-\mathbf{u}_j^{(l)}\|_2^2 .
\end{equation}
Only the final $L_2$-regularized logistic classifier operating on the 200 Gaussian features used labels; it employed the same optimization parameters as the linear probe. The model therefore tests a fixed-capacity, locally nonlinear notion of class accessibility without label-supervised adjustment of centers or bandwidths.

\paragraph{RBF support-vector machine.}
\NI As a stronger nonlinear reference, an exact support-vector classifier with Gaussian radial-basis kernel
\begin{equation}
K(\mathbf{z},\mathbf{z}')=
\exp\!\left[-\gamma^{(l)}\|\mathbf{z}-\mathbf{z}'\|_2^2\right]
\end{equation}
was trained on the same fixed, class-stratified subset of 10,000 training examples from the respective dataset for every layer and every DBN. The input dimensionality of the representation is denoted by $d_l$. We used $C=10$ and $\gamma^{(l)}=4/d_l$. These values had been selected in a preliminary MNIST training-only validation using a common parameter grid, $C\in\{1,10\}$ and $\gamma d_l\in\{0.25,1,4\}$, evaluated jointly at H1, H5, and H10 on an 8,000/2,000 training-validation split. They were subsequently frozen for all five main-study seeds and all three datasets. Because the RBF-SVM used 10,000 rather than 60,000 training examples, its absolute accuracy should not be interpreted as a capacity-matched comparison with the other probes; its layer-wise trend is the primary quantity of interest.

\subsection{H1 weight-permutation control}

\NI To test whether changes in class accessibility from the input to the first hidden layer depended on learned feature structure rather than merely on the marginal distribution of weight values, we constructed a structurally destroyed H1 control for MNIST, Fashion-MNIST, and KMNIST. For every trained DBN and every H1 unit $j$, the 784 entries of its incoming weight vector were independently permuted across visible coordinates,
\begin{equation}
\widetilde{W}_{ij}=W_{\pi_j(i),j},
\end{equation}
where $\pi_j$ was an independent random permutation for unit $j$. This operation preserved exactly the multiset, mean, variance, and Euclidean norm of the incoming weights of each neuron, as well as its hidden bias, while destroying the learned association between individual pixel positions and weight values. Permutation seeds were deterministically coupled to the corresponding DBN seeds. Whole hidden units were not permuted, since such a permutation would only reorder representation coordinates and leave probe performance unchanged.

\NI For each dataset, deterministic activation probabilities were computed for the intact and permuted H1 representations of the same five 10-by-100 CD-1 models used in Fig.~\ref{Fig1}. Linear, Gaussian-kernel, and RBF-SVM probes were then refitted from scratch on the permuted representations using exactly the protocols described above. The comparison comprised the original input, intact learned H1, and permuted H1. Thus, any performance difference between intact and permuted H1 cannot be attributed to neuron count, per-neuron weight distributions, weight norms, or biases, but requires the learned assignment of weights to input coordinates. Because permutation may also alter activation distributions and saturation, this control establishes the necessity of learned weight structure but does not by itself identify which aspect of that structure causes the performance change.

\NI To separate the effects of the learned linear projection and the sigmoid nonlinearity, we performed an additional exploratory MNIST analysis on the seed-1234 model. For the intact and permuted H1 weights, probes were evaluated both on the preactivations $\mathbf{a}=\mathbf{x}W+\mathbf{b}$ and on the activation probabilities $\sigma(\mathbf{a})$. We further constructed a continuous series of weight matrices,
\begin{equation}
W_{\alpha}=\alpha W+(1-\alpha)\widetilde{W}, \qquad
\alpha\in\{0,0.25,0.5,0.75,1\},
\end{equation}
and rescaled every column of $W_{\alpha}$ to the Euclidean norm of the corresponding learned weight vector. This interpolation varied the retained learned pixel structure while holding every neuron's weight norm fixed. For each $\alpha$, the linear and RBF-SVM probes were fitted using the protocols above. As simple activation-distribution diagnostics, we calculated the mean standard deviation of the H1 preactivation across neurons and the fraction of sigmoid activations below 0.01 or above 0.99. The latter was defined as the saturation fraction. Because this mechanism analysis used a single pretrained DBN, it is interpreted as a targeted exploratory control rather than a multi-seed estimate.

\subsection{Pairwise generalized discrimination value}

\NI To resolve the global GDV into individual class relations, a pairwise GDV was calculated for each of the 45 unordered MNIST digit pairs. Within a given layer, standardization was performed once over the complete fixed, class-balanced evaluation subset using the same transformation as for the global GDV,
\begin{equation}
z_{n,d}^{(l)}=
\frac{h_{n,d}^{(l)}-\mu_d^{(l)}}{2\sigma_d^{(l)}} .
\end{equation}
Dimensions with zero variance were set to zero. For class $a$ with $N_a$ samples, the mean within-class distance was
\begin{equation}
d_{\mathrm{intra},a}^{(l)}=
\frac{2}{N_a(N_a-1)\sqrt{D_l}}
\sum_{\substack{i<j\\y_i=y_j=a}}
\left\|\mathbf{z}_i^{(l)}-\mathbf{z}_j^{(l)}\right\|_2 ,
\end{equation}
and the mean distance between classes $a$ and $b$ was
\begin{equation}
d_{\mathrm{inter},ab}^{(l)}=
\frac{1}{N_aN_b\sqrt{D_l}}
\sum_{\substack{i:y_i=a\\j:y_j=b}}
\left\|\mathbf{z}_i^{(l)}-\mathbf{z}_j^{(l)}\right\|_2 .
\end{equation}
The pairwise GDV was then defined as
\begin{equation}
G_{ab}^{(l)}=
\frac{1}{2}\left(d_{\mathrm{intra},a}^{(l)}+
d_{\mathrm{intra},b}^{(l)}\right)
-d_{\mathrm{inter},ab}^{(l)},
\qquad a\neq b .
\end{equation}
As for the global GDV, more negative values indicate better geometric separation, whereas an increase toward zero indicates pairwise deterioration. The matrix is symmetric and its diagonal is undefined. Pairwise matrices were evaluated using 200 fixed test examples per digit and the same evaluation subset for every DBN, then averaged across five DBN seeds. With equally sized classes, the mean of the 45 pairwise values equals the corresponding global GDV.

\NI To relate geometric changes to classification errors, the RBF-SVM confusion matrix $\mathbf{M}^{(l)}$ was converted into a symmetric confusion rate for each pair,
\begin{equation}
C_{ab}^{(l)}=
\frac{M_{ab}^{(l)}+M_{ba}^{(l)}}{N_a^{\mathrm{test}}+N_b^{\mathrm{test}}} .
\end{equation}
Changes were defined relative to H2 as $\Delta G_{ab}=G_{ab}^{(10)}-G_{ab}^{(2)}$ and $\Delta C_{ab}=C_{ab}^{(10)}-C_{ab}^{(2)}$. Thus, positive values denote deterioration in both cases. In Fig.~\ref{Fig4}, a pair was outlined only when both $\Delta G_{ab}>0$ and the seed-averaged $\Delta C_{ab}>0$.

\section{Results}

\subsection{Layer-wise evolution of class separation, abstraction distance and effective dimensionality}

\NI Our first goal is to train deep belief networks (DBNs) on different image datasets in a completely unsupervised manner, without using any class labels. Training is performed using the standard contrastive-divergence method with one Gibbs step (CD-1). The DBNs are trained greedily, layer by layer, with one restricted Boltzmann machine (RBM) being trained after another. No subsequent fine-tuning of the complete network is applied.

\NI To facilitate comparison, we use three datasets that share the same format but differ in their visual content and complexity: MNIST, containing handwritten digits; Fashion-MNIST, containing images of clothing items; and KMNIST, containing handwritten Japanese Kuzushiji characters. Each dataset consists of grayscale images with
$28 \times 28 = 784$
pixels and contains $60{,}000$ training images and $10{,}000$ test images.

\NI Accordingly, our DBNs have $784$ input units in the visible layer, followed by ten stacked hidden layers, each containing the same number $N_{\mathrm{H}}$ of neurons. We compare three hidden-layer sizes:
$N_{\mathrm{H}} \in \{100, 50, 30\}$.

\NI After training, we present a fixed, class-balanced subset of test images to the visible layer of each DBN and propagate the resulting activity upward through the network hierarchy. For every input image, we record the activation probabilities of all neuronal units in all hidden layers. Based on these input-driven activation probabilities, we then calculate three aggregate quantities for each layer. These quantities characterize the emergence of cluster-like class separation, the abstraction distance between inputs and their reconstructions, and the effective degree of data compression achieved through dimensionality reduction.

\paragraph{First, we investigate class separation.}
In each layer of the DBN, the test dataset is represented as a point cloud in a high-dimensional space. We investigate the extent to which the ten different classes form distinct subclusters within this point cloud. Two-dimensional visualizations of these representations are shown in Fig.~\ref{Fig2}(a--e).

\NI To quantify the degree of class-specific clustering, we compute the Generalized Discrimination Value (GDV), defined as a normalized difference between the mean Euclidean intra-class distance and the mean inter-class distance. A detailed description of its invariant and robust properties is provided in the Methods section and in several of our previous publications \cite{schilling2021quantifying}. According to the definition of the GDV, more negative values indicate stronger class separation. For approximately Gaussian class distributions, values close to $-1$ correspond to extremely well-separated class-specific clusters.

\begin{figure}[p]
\centering
\includegraphics[width=\textwidth]{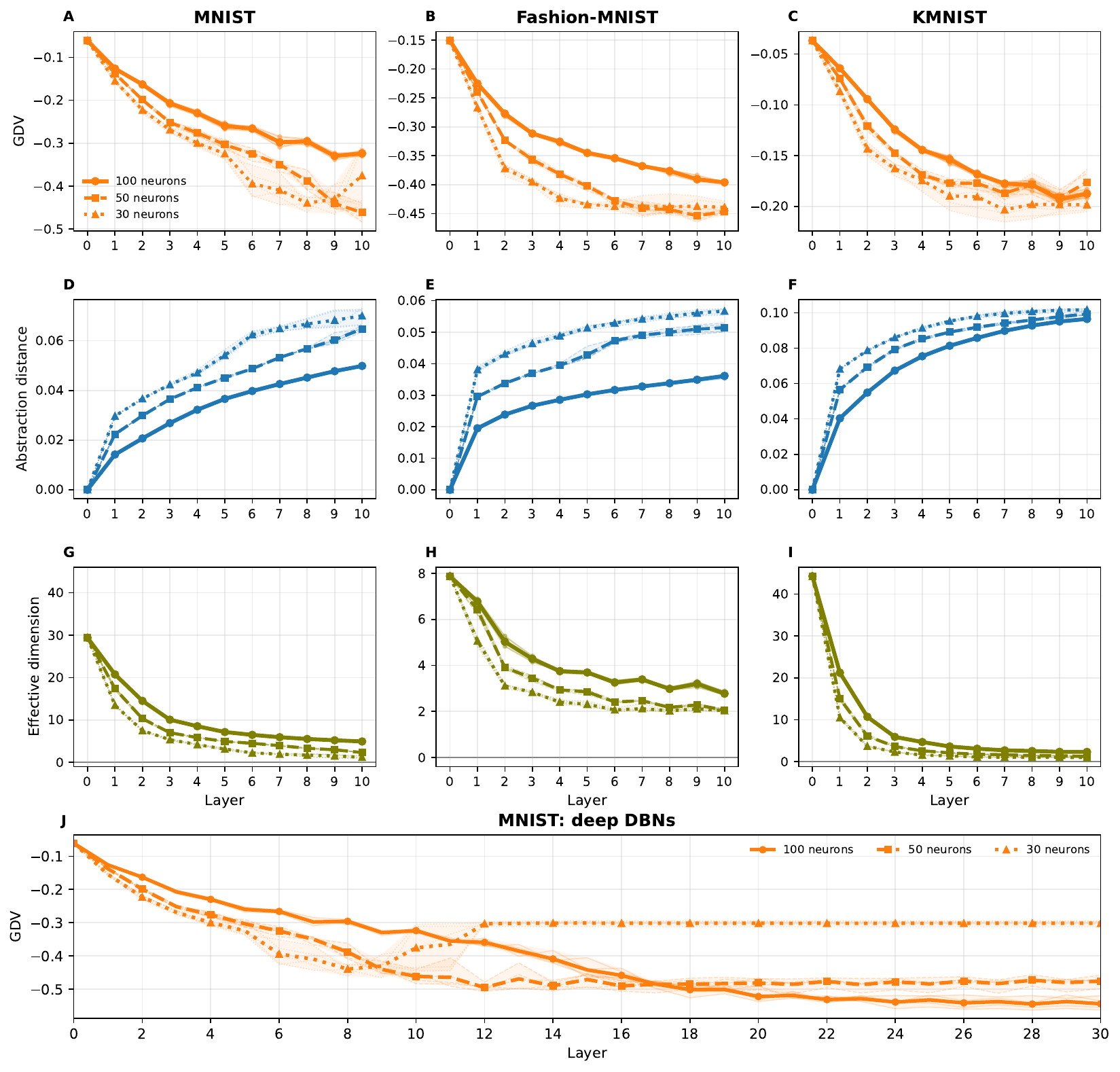}
\caption{{\bf Layer-wise evolution of representation measures across three image-classification tasks.}
Deep belief networks with 100, 50, or 30 neurons per hidden layer were trained using CD-1 on MNIST, Fashion-MNIST, and KMNIST. Solid lines with circles, dashed lines with squares, and dotted lines with triangles denote widths of 100, 50, and 30 neurons, respectively. 
\textbf{(A--I)} The generalized discrimination value (GDV; orange), abstraction distance operationalized as reconstruction MSE (blue), and effective dimension (olive) are shown as functions of layer depth for ten-layer DBNs. 
\textbf{(J)} MNIST GDV for matched, exceptionally deep DBNs comprising 30 hidden layers. The ten lower RBMs are identical to those used in panels A, D, and G; layers H11--H30 were subsequently trained with the same greedy CD-1 procedure and hyperparameters while the lower layers remained frozen. Thin curves represent five individual random seeds, shaded regions indicate the corresponding minimum--maximum ranges, and thick curves show the means across seeds. GDV was evaluated using the same fixed, class-balanced subset for all models.}
\label{Fig1}
\end{figure}

\NI For the DBN with $N_{\mathrm{H}}=100$ neurons per hidden layer, the GDV decreases almost monotonically across successive layers for all three investigated datasets, as shown by the solid orange lines in Fig.~\ref{Fig1}(A--C). The relative improvement in the GDV between the input layer and the highest hidden layer is greater for the simpler MNIST and Fashion-MNIST datasets than for the more complex KMNIST dataset. The decrease in the GDV becomes even more pronounced when the hidden-layer width $N_{\mathrm{H}}$ is reduced, as indicated by the orange dashed and dotted lines.

\NI These results indicate that, across successive layers of the DBN, the different input classes form progressively better-separated clusters, even though no class labels were provided during unsupervised training.

\NI For the MNIST dataset, we additionally investigated a substantially deeper DBN comprising 30 stacked hidden layers (Fig.~\ref{Fig1}(J)). In the narrow network with $N_{\mathrm{H}}=30$ neurons per hidden layer, shown by the dotted curve, the GDV reaches a minimum at hidden layer 8, subsequently increases until approximately layer 12, and then remains nearly constant. In the medium-width network with $N_{\mathrm{H}}=50$, shown by the dashed curve, the GDV decreases until about layer 12 and thereafter fluctuates around an approximately constant level. In the wide network with $N_{\mathrm{H}}=100$, shown by the solid curve, the GDV continues to decrease over a much larger number of layers and reaches a low plateau only at around layer 24.

\NI The minimum observed in the narrow network suggests a qualitative change in the evolution of the internal data representation near layer 8. Beyond this point, the previous trend toward improved class separation is reversed, and the class-specific subclusters begin to overlap more strongly. After approximately layer 12, all subsequent layers appear to produce essentially the same collapsed representation of the data.

\paragraph{Second, we investigate abstraction distance.}
For this purpose, we present a test pattern $x$ to the input layer of the DBN, propagate the continuous activation probabilities upward to a specific hidden layer $n$, and then propagate them downward again to the input layer. This procedure yields a reconstructed input pattern $\hat{x}$. Repeating the process for all test patterns, we compute the mean squared input--reconstruction difference shown in Fig.~\ref{Fig1}(D--F). We refer to this general quantity as the abstraction distance, while retaining reconstruction MSE as its mathematical operationalization. By definition, the abstraction distance is zero at the input layer.

\NI We find that the abstraction distance increases monotonically across successive DBN layers. This increase becomes more pronounced as the width of the hidden layers decreases and is largest for the more complex KMNIST dataset. The scalar increase alone does not determine whether the transformation is beneficial: it may reflect a prototypical shift, destructive information loss, or a mixture of both.

\NI These results indicate that higher layers of a DBN produce reconstructions at progressively greater abstraction distance from the original input pattern. Inspection of individual examples in Fig.~\ref{Fig6} shows that the early increase is mainly associated with the omission or alteration of small, often idiosyncratic details. In this regime, the transformation can be described as a prototypical shift toward simplified, class-typical forms. In the higher layers, however, the shift increasingly includes distortions that may remove class-discriminative details or change the apparent class. Abstraction and degradation therefore become progressively intertwined.

\paragraph{Finally, we investigate the effective dimensionality.}
Although a hidden layer contains a prescribed number $N_{\mathrm{H}}$ of neurons, it is not guaranteed that all of them learn distinct and useful features. Some neurons may become effectively frozen and exhibit only negligible activity variance, whereas others may acquire highly redundant feature representations. In either case, the effective dimensionality of the representation is smaller than the nominal layer width $N_{\mathrm{H}}$.

\NI We quantify this dimensionality reduction separately for each layer using the participation ratio of the covariance spectrum, computed from the continuous neuronal activation probabilities. The resulting effective dimension, described in detail in the Methods section, approaches one when nearly all variance is concentrated along a single direction. Conversely, it approaches the algebraic rank of the covariance matrix when the variance is distributed approximately uniformly across all active directions.

\NI We find that the effective dimensionality decreases monotonically across successive DBN layers, with a more pronounced reduction in networks with narrower hidden layers (Fig.~\ref{Fig1}(G--I)). In some cases, the effective dimension approaches one in the final layers of the stack. Relative to the effective dimensionality of the input representation, this corresponds to an approximately 30-fold dimensionality collapse for the narrow network in the MNIST experiment.

\NI These results indicate that DBNs can produce an extreme reduction in effective dimensionality. This reduction is not imposed directly by the prescribed layer width, but instead emerges from the layerwise contrastive-divergence training procedure.

\subsection{Control experiments for unsupervised class separation}

\NI In our previous investigation of the GDV, we found that even transformations without any task-specific learning can cause small changes in class separation \cite{schilling2021quantifying}. In particular, random linear transformations slightly broaden the distribution of possible GDV changes and, on average, tend to weakly reduce rather than improve class separation. Adding a sigmoid nonlinearity produces another small shift. A more pronounced artificial decrease in the GDV can occur when the inputs to sigmoidal neurons are scaled so strongly that the neurons are driven into saturation. By contrast, changes in the number of representation dimensions do not systematically change the GDV because of its dimensionality normalization. These observations motivate additional controls to distinguish class-specific representational changes induced by learning from generic GDV shifts caused by linear transformations, nonlinearities, or neuronal saturation. We therefore performed a series of control experiments to determine which components of the DBN are responsible for this effect.

\begin{figure}[p]
\centering
\includegraphics[width=\textwidth]{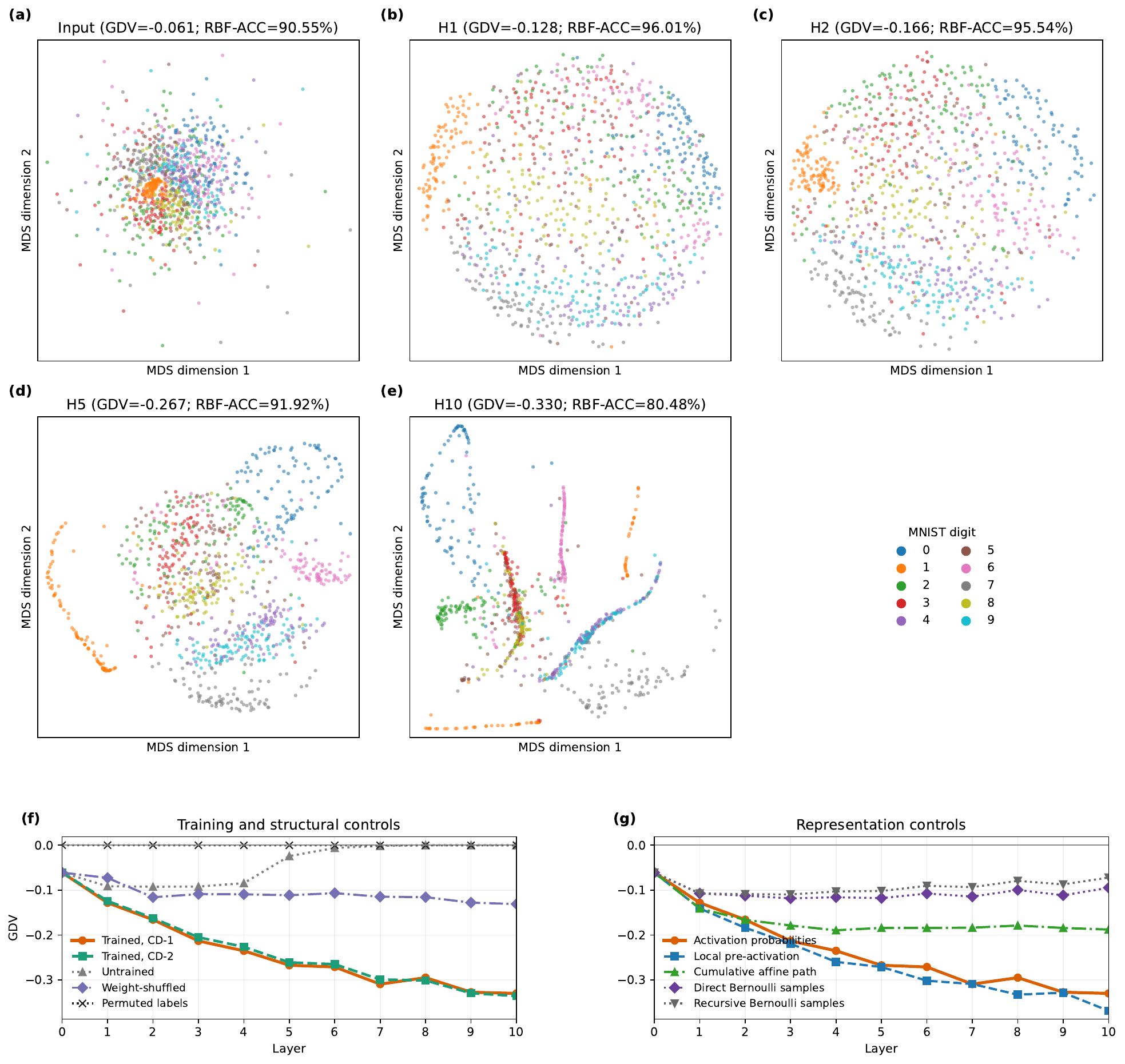}
\caption{{\bf Layer-wise transformation of MNIST representations and control analyses.}
\textbf{(a--e)} Metric two-dimensional multidimensional-scaling (MDS) embeddings of the input and the deterministic activation-probability representations in hidden layers H1, H2, H5, and H10 of a DBN trained with CD-1 (seed 1234). The same class-balanced subset of 100 test examples per MNIST digit was used in every panel; colors denote digit labels. Distances were computed after the same component-wise standardization used for the GDV. Panel titles report the GDV and the test accuracy of an RBF-SVM trained on the respective full representation. MDS serves only as a visualization and was not involved in DBN training or quantitative GDV estimation. \textbf{(f)} Layer-wise GDV for the trained CD-1 reference, a matched CD-2 run, an untrained network, a network with independently shuffled incoming weight vectors, and the trained representations evaluated after permutation of the labels. \textbf{(g)} Comparison of deterministic activation probabilities with local pre-activations, the cumulative affine control path, and direct or recursively propagated Bernoulli samples. All GDV curves show the same MNIST seed-1234 preliminary experiments.}
\label{Fig2}
\end{figure}

\NI We first compared the standard CD-1 network with an otherwise matched DBN trained using two Gibbs steps per update (CD-2). Both networks produced almost identical layer-wise GDV curves (Fig.~\ref{Fig2}(f)). The progressive class separation is therefore not a peculiarity of the one-step approximation used for contrastive-divergence training.

\NI Next, we propagated the MNIST images through a network with the same architecture but randomly initialized, untrained weights. This random network produced only a small initial decrease in GDV, from approximately $-0.06$ in the input to $-0.09$ in the first hidden layers. In the subsequent layers, the absolute activity variance rapidly vanished and the GDV approached zero. Thus, repeated random affine transformations followed by sigmoid nonlinearities do not reproduce the strong and sustained decrease observed in the trained network.

\NI To separate the contribution of the learned weight values from that of their learned arrangement, we independently permuted the entries of every incoming weight vector in the trained DBN. This operation preserved the complete weight distribution, mean, norm, and bias of every neuron, but destroyed the association between individual input coordinates and their corresponding weights. The resulting GDV remained close to approximately $-0.1$ across most layers and reached only about $-0.13$ in H10, compared with approximately $-0.33$ for the intact network (Fig.~\ref{Fig2}(f)). Consequently, the marginal weight statistics alone are insufficient. Most of the observed class separation depends on the specific feature structure learned by contrastive divergence.

\NI As an additional negative control, we retained the trained representations but randomly permuted the digit labels used for calculating the GDV. The GDV then remained close to zero in every layer. Hence, the falling GDV does not reflect arbitrary subdivisions of the point cloud or an unspecific tendency to form ten clusters. It specifically describes the increasing geometric separation of the actual MNIST digit classes, although these labels were not available during DBN training.

\NI We then investigated whether the decrease in GDV was produced mainly by the sigmoid activation function (Fig.~\ref{Fig2}(g)). For every layer, we compared the GDV immediately before and after application of the local sigmoid. In most layers, the pre-activation GDV was slightly more negative than the corresponding post-sigmoid value. The sigmoid therefore generally reduced some of the separation already generated by the weighted affine transformation instead of creating it. This result is particularly important because many units operated partly in the saturated regime. Saturation can affect the precise GDV value, but it cannot explain the progressive decrease across the trained DBN.

\NI In a complementary control, we propagated the input cumulatively through all trained weights and biases while omitting every sigmoid nonlinearity. Along this purely affine path, the GDV initially decreased but reached a plateau near $-0.18$, substantially above the value obtained with the complete DBN. The learned linear transformations therefore contain class-relevant structure, but the full layer-wise improvement requires their repeated interaction with the nonlinear representations supplied to the subsequent RBMs. Neither the affine transformations nor the sigmoid nonlinearities alone account for the complete effect.

\NI Finally, the preceding analyses used deterministic activation probabilities, which describe the conditional mean representation of each hidden layer. We therefore repeated the GDV analysis with Bernoulli states sampled from these probabilities. Direct sampling strongly weakened the separation, and recursively feeding each sampled state into the next RBM weakened it further. Nevertheless, the first hidden layer still showed a more negative GDV than the input. These results demonstrate that stochastic sampling adds substantial representational noise but does not generate the decrease observed for the activation probabilities. They also clarify that the strong layer-wise GDV effect characterizes the probability-coded representation of the DBN and is only partially preserved in individual binary state realizations.

\NI Taken together, the controls show that the unsupervised increase in class separation is a genuine consequence of learning. It is robust to the number of Gibbs steps used for contrastive divergence, disappears for random labels, is much weaker in untrained or structurally shuffled networks, and cannot be explained by dimensionality changes or sigmoid saturation. The learned affine weight structure produces an important part of the effect, while the successive nonlinear representations enable its continued accumulation across layers. Stochastic binary sampling counteracts this organization by adding noise. Because the controls in Fig.~\ref{Fig2}(f,g) were performed as preliminary experiments for one MNIST seed, their quantitative magnitudes should not yet be interpreted as population estimates; their mutually consistent qualitative pattern, however, rules out the principal trivial explanations for the falling GDV.

\subsection{Origin of the dimensional collapse}

\begin{figure}[p]
\centering
\includegraphics[width=\textwidth]{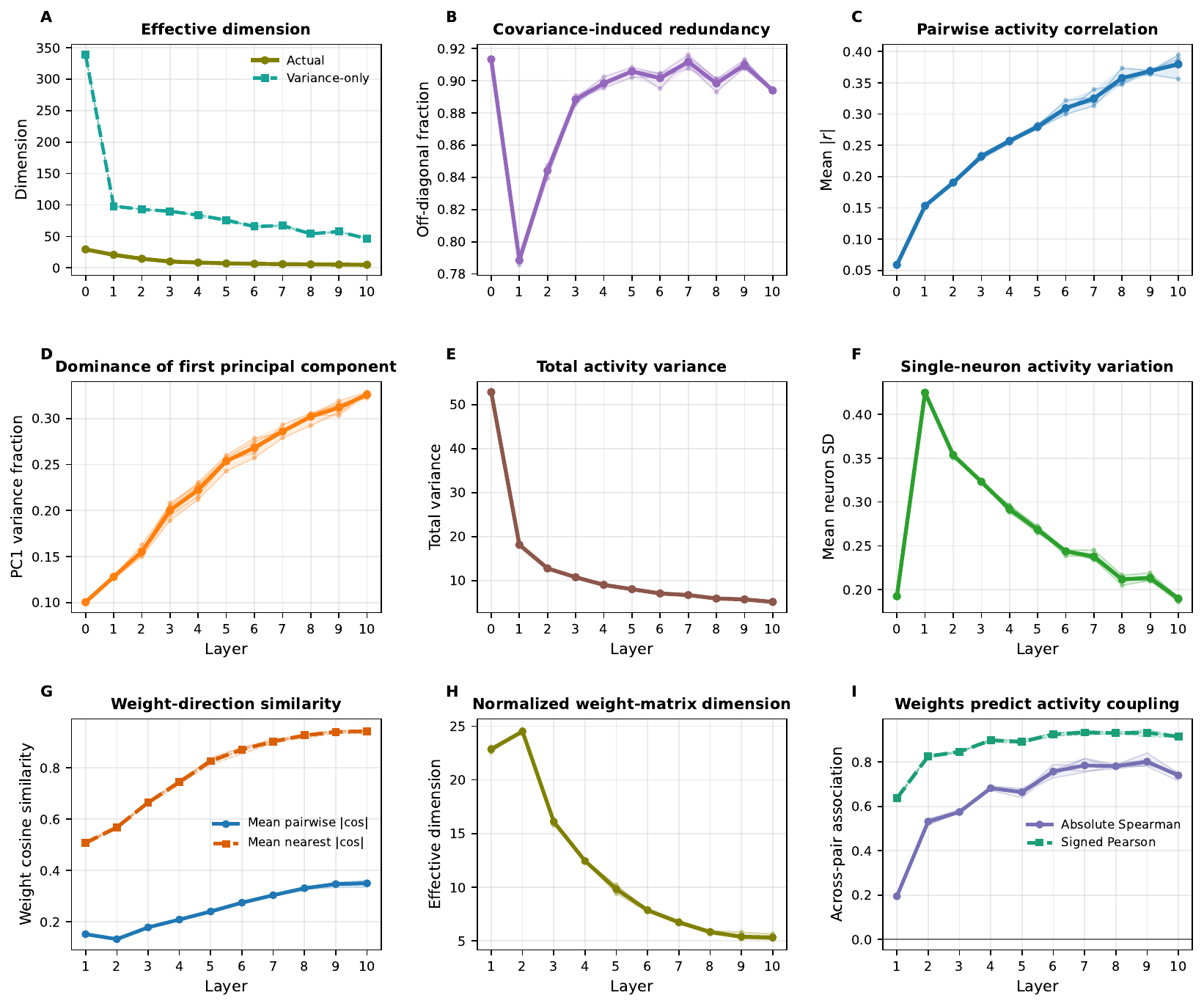}
\caption{{\bf Redundancy of learned weight directions drives the layer-wise collapse of effective dimensionality.}
Five MNIST DBNs with 100 neurons per hidden layer were trained using CD-1. Thin curves show individual random seeds, shaded regions indicate the corresponding minimum--maximum ranges, and thick curves show means across seeds; layer 0 denotes the input representation and is included only in the activity-based panels A--F. \textbf{(A)} Actual effective dimensionality (olive) and the variance-only dimension obtained after discarding all off-diagonal covariances (turquoise). Their large separation shows that unequal single-neuron variances alone cannot account for the observed dimensionality collapse. \textbf{(B)} Covariance-induced redundancy, defined as $1-D_{\mathrm{eff}}/D_{\mathrm{var}}$, equivalently the fraction of squared covariance-matrix energy contained in off-diagonal entries. \textbf{(C)} Mean absolute Pearson correlation between pairs of non-constant neuronal activities. \textbf{(D)} Fraction of total variance explained by the first principal component. \textbf{(E)} Total activity variance, given by the trace of the covariance matrix. \textbf{(F)} Mean single-neuron activity standard deviation. \textbf{(G)} Mean absolute cosine similarity across all incoming weight-vector pairs and mean similarity of each neuron to its most similar same-layer partner. \textbf{(H)} Participation-ratio dimension of the weight matrix after normalizing every neuronal weight vector to unit length. \textbf{(I)} Across-pair association between weight-vector similarity and activity correlation, shown as the Spearman correlation between their absolute values and the Pearson correlation between their signed values. Together, the measures show that deeper representations become dominated by a small number of shared activity directions because many neurons learn increasingly similar weight directions and consequently produce correlated, functionally redundant responses.}
\label{Fig3}
\end{figure}

\NI Our next goal is to determine what causes the pronounced decline in effective dimensionality observed across successive DBN layers in Fig.~\ref{Fig1}(G--I). A low effective dimension can arise for at least two fundamentally different reasons. First, most neurons could become nearly constant, leaving only a small number of variable units. Second, many neurons could remain responsive but vary in a coordinated manner, so that their population activity occupies only a few common directions. To distinguish these possibilities, we decomposed the covariance structure of the layer-wise MNIST representations using the six complementary measures shown in Fig.~\ref{Fig3}.

\NI We first compared the actual effective dimension with a variance-only control in which all covariances between different neurons were set to zero while the variance of every individual neuron was retained (Fig.~\ref{Fig3}(A)). From H1 to H10, the actual effective dimension decreases from approximately 20.7 to 4.9. The variance-only dimension also decreases, showing that the distribution of activity variance across neurons becomes less uniform. However, it remains as high as approximately 46 in H10. Moreover, all 100 neurons remain variable according to the numerical activity criterion used in the analysis. Unequal single-neuron variances and inactive units therefore contribute to the decline, but cannot explain why the population behaves as an effectively five-dimensional system.

\NI The discrepancy between the two dimensions is quantified directly by the covariance-induced redundancy shown in Fig.~\ref{Fig3}(B). This quantity measures the fraction of squared covariance-matrix energy contained in off-diagonal entries. It is already approximately 0.79 in H1, increases to about 0.91 in the intermediate layers, and remains close to 0.9 in the upper layers. Thus, the dominant part of the covariance structure is generated by dependencies between different neurons rather than by their individual variances. The pixel representation in layer 0 also has a large off-diagonal fraction because neighboring and anatomically related pixels are naturally dependent. For the present question, the important observation is the consistently high redundancy within the hidden-layer representations.

\NI To obtain a more intuitive measure of these dependencies, we calculated the mean absolute Pearson correlation between the activity patterns of all neuron pairs (Fig.~\ref{Fig3}(C)). For each neuron, its activity pattern is the sequence of responses elicited by the 10,000 MNIST test images. The mean absolute correlation increases almost monotonically from approximately 0.15 in H1 to 0.38 in H10. Positive and negative correlations both contribute because either form of coupling reduces the independence of the neuronal responses. Hence, neurons in deeper layers respond to different input images in an increasingly coordinated manner.

\NI The same change can be viewed from the eigenspectrum of the covariance matrix. Figure~\ref{Fig3}(D) shows the fraction of total activity variance explained by the first principal component. This fraction increases from approximately 13\% in H1 to 33\% in H10. Consequently, a single shared population mode accounts for about one third of all activity variance in the highest hidden layer. The progressive dominance of this first principal component confirms that the representation becomes concentrated along a small number of common activity directions.

\NI In parallel with this reorganization, the overall size of the activity point cloud decreases (Fig.~\ref{Fig3}(E)). The total activity variance falls from approximately 18.1 in H1 to 5.2 in H10. The hidden-layer representations therefore become increasingly compact as network depth increases. However, a uniform contraction of all activity directions would leave the participation ratio, and thus the effective dimension, unchanged. The decline in total variance accompanies the dimensional collapse but is not by itself its cause.

\NI Finally, we measured the average activity variation of individual neurons (Fig.~\ref{Fig3}(F)). The mean single-neuron standard deviation initially increases from approximately 0.19 in the input representation to 0.42 in H1 and subsequently decreases to approximately 0.19 in H10. Individual neuronal responses therefore become weaker in the higher layers, consistent with the contraction observed in panel E. Nevertheless, the neurons do not generally become constant, and their remaining variance would still support a variance-only dimension of approximately 46. Reduced single-neuron variation is therefore again insufficient to account for the observed effective dimension of approximately five.

\NI The preceding activity measures establish functional redundancy, but they do not determine whether this redundancy is already present in the features learned by the individual neurons. We therefore directly compared the incoming weight vectors of all neuron pairs within each RBM layer. Because all neurons in a given layer receive input from the same preceding representation, their weight vectors can be compared without an alignment or neuron-permutation problem.

\NI We first calculated the absolute cosine similarity between the incoming weight vectors (Fig.~\ref{Fig3}(G)). The mean similarity across all neuron pairs initially decreases slightly from approximately 0.15 in H1 to 0.13 in H2, but subsequently rises almost monotonically to 0.35 in H10. An even clearer result is obtained by identifying the most similar partner of each neuron. The mean nearest-neighbour similarity increases from approximately 0.51 in H1 to 0.94 in H10. Thus, a typical neuron in the highest layer has at least one other neuron with an almost parallel or antiparallel incoming weight vector. This result directly demonstrates the emergence of near-duplicate feature directions in the deeper RBMs.

\NI To characterize the complete set of learned weight directions, we next calculated the participation-ratio dimension of each weight matrix after normalizing every neuronal weight vector to unit length (Fig.~\ref{Fig3}(H)). This normalization ensures that the result reflects directional diversity rather than differences in weight norm. The effective weight dimension is approximately 22.9 in H1, briefly increases to 24.5 in H2, and then collapses almost monotonically to 5.3 in H10. Remarkably, this final value is close to the effective dimension of approximately 4.9 measured for the H10 activity representation. The learned features and the resulting neuronal activities therefore collapse onto similarly small subspaces.

\NI Finally, we tested whether similar weight vectors actually predict similar neuronal responses (Fig.~\ref{Fig3}(I)). Across all neuron pairs, the Pearson correlation between signed weight cosine similarity and signed activity correlation increases from approximately 0.64 in H1 to 0.91 in H10. Parallel weight vectors therefore tend to produce positively correlated responses, whereas antiparallel directions tend to produce negatively correlated responses. Consistently, the Spearman association between absolute weight similarity and absolute activity correlation increases from approximately 0.19 to 0.74. Weight direction thus becomes an increasingly strong predictor of functional coupling in the upper DBN layers.

\NI Taken together, the nine measures identify learned feature redundancy as the main origin of the dimensional collapse. With increasing layer depth, the representation contracts and the distribution of variance across neurons becomes less uniform, but these scale-related changes explain only part of the effect. The decisive change is that many still-variable neurons learn increasingly aligned incoming weight directions and consequently produce statistically dependent responses. The weight matrix itself collapses to nearly the same effective dimension as the evoked population activity. Deeper DBN layers therefore do not primarily reduce dimensionality by switching neurons off. Instead, the local unsupervised learning process repeatedly produces overlapping feature directions, so that many nominally distinct neurons carry closely related information.

\subsection{GDV and actual class accessibility}

\NI The GDV is particularly useful for quantifying class-specific clustering in high-dimensional point clouds when the individual class distributions form approximately compact, spherical clusters. It summarizes the geometry of the complete representation by averaging the distances between points from the same class and comparing them with the distances between points from different classes. This provides a parameter-free and non-invasive measure of overall class separation, but it does not construct a decision boundary. Consequently, two representations with similar GDV values need not be equally useful to a particular classifier. Even compact class clusters may be arranged or overlap in such a way that they cannot all be separated equally well by linear decision surfaces, while a more flexible nonlinear classifier may exploit additional geometric structure. Conversely, a favorable average over all inter-class distances can coexist with a small number of local class overlaps that dominate the actual classification errors.

\begin{figure}[p]
\centering
\includegraphics[width=\textwidth,height=0.70\textheight,keepaspectratio]{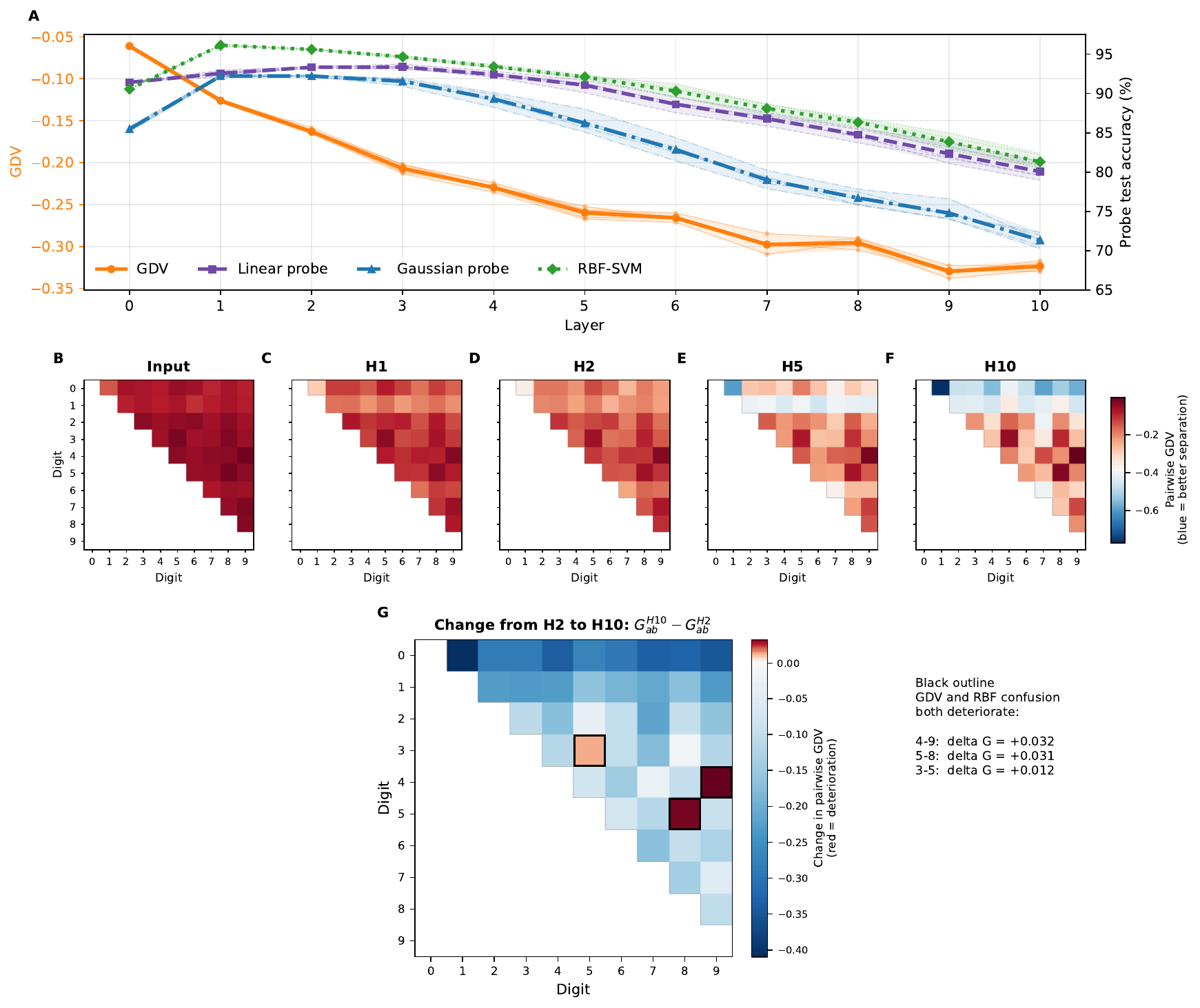}
\caption{{\bf Global GDV improvement coexists with declining class readout and the collapse of a few critical digit pairs.}
Five MNIST DBNs with 100 neurons per hidden layer were trained using CD-1; layer 0 denotes the input representation. \textbf{(A)} Layer-wise GDV and test accuracy of three probes fitted independently to each frozen representation. Thin curves show individual DBN seeds, shaded regions indicate minimum--maximum ranges, and thick curves show means. Linear logistic and 200-center Gaussian-kernel probes were trained on all 60,000 training examples. Gaussian centers were obtained without labels by MiniBatchKMeans, with a shared layer-specific bandwidth set to the median training distance to the nearest center; only the logistic output used labels. The exact RBF-SVM used the same fixed, stratified subset of 10,000 training examples for every DBN, with $C=10$ and $\gamma=4/d$ fixed from the preliminary training-only validation. Probe levels should therefore be compared cautiously across methods, whereas their layer-wise trends are directly interpretable. \textbf{(B--F)} Mean pairwise-GDV matrices for the input, H1, H2, H5, and H10, evaluated on the same fixed class-balanced test subset and averaged across five DBN seeds. Only the upper triangle is shown because the matrices are symmetric; all panels use a common color scale, with more negative values indicating better separation. \textbf{(G)} Change in pairwise GDV from H2 to H10. Blue cells denote improved and red cells deteriorated pairwise separation. Black outlines identify pairs for which both pairwise GDV and symmetric RBF-SVM confusion deteriorated: 4--9, 5--8, and 3--5. Thus, the global GDV improves because most digit pairs become geometrically better separated, while the decline in classification accuracy is concentrated in a small number of difficult, visually similar pairs.}
\label{Fig4}
\end{figure}

\NI It is therefore important to assess not only the geometric clustering measured by the GDV, but also how much class information can actually be extracted from each frozen DBN layer. For this purpose, we trained three supervised probe models independently on the input representation and on every hidden-layer representation (Fig.~\ref{Fig4}(A)). The DBN itself remained unchanged, and class labels were used only to fit and evaluate the probes. We used a linear logistic probe, a finite Gaussian-kernel probe with 200 unlabeled cluster centers, and an exact nonlinear RBF support-vector machine. These models provide complementary readouts with different decision-boundary flexibility.

\NI The linear probe already achieves a test accuracy of approximately 91.4\% on the original MNIST pixels. Its accuracy increases to 92.6\% in H1 and reaches a maximum of approximately 93.3\% in H2. Thus, the first two unsupervised DBN layers make digit identity more accessible even to a linear readout. Beyond H2, however, the accuracy decreases continuously and reaches only approximately 80.1\% in H10. The early layers therefore improve linear accessibility, whereas the deeper layers progressively remove information required for reliable linear classification.

\NI The two nonlinear probes show the same qualitative pattern. The Gaussian-kernel probe improves from approximately 85.5\% on the input to 92.2\% in H1 and H2, but subsequently declines to 71.3\% in H10. The exact RBF-SVM increases from approximately 90.6\% on the input to 96.1\% in H1, followed by a gradual decline to 81.3\% in H10. The absolute accuracies of the three probes should not be compared as a capacity-matched ranking because their training-set sizes and nonlinear parameterizations differ. Their common layer-wise trend is nevertheless robust: class accessibility improves in the first DBN layer, remains high or reaches its maximum in H2, and then deteriorates substantially across the higher layers.

\NI This behavior differs fundamentally from the GDV curve. Across the same representations, the mean GDV decreases from approximately $-0.061$ in the input to $-0.324$ in H10 and therefore indicates progressively stronger overall class clustering. The agreement between three substantially different probe models makes it unlikely that their common decrease merely reflects an unsuitable decision-boundary family. Instead, GDV and probe accuracy quantify related but non-identical properties. The GDV describes the average geometry of labeled point clouds, whereas probe accuracy additionally depends on the local arrangement of decision boundaries, the preservation of discriminative details, and the particular class pairs that remain ambiguous.

\NI To determine whether this discrepancy is concentrated in specific digit pairs, we decomposed the global GDV into all 45 pairwise contributions (Fig.~\ref{Fig4}(B--F)). Each matrix element compares the mean within-class spread of two digits with the mean distance between those two classes. The matrices therefore reveal which class pairs contribute to the increasingly negative global GDV and whether a favorable global average conceals local deterioration.

\NI From the input through H1, H2, H5, and H10, most matrix elements become progressively more negative, producing the visually dominant blue structure in the higher-layer matrices. A direct comparison between H2 and H10 confirms that 42 of the 45 digit pairs improve according to their pairwise GDV (Fig.~\ref{Fig4}(G)). This widespread improvement explains why the global GDV continues to decrease even while every probe loses accuracy. The geometric reorganization is real, but it primarily increases the average separation of the majority of class pairs.

\NI Three pairs form a notable exception: 4--9, 5--8, and 3--5 show an increase in pairwise GDV from H2 to H10 and therefore become geometrically less separated. Importantly, these are also the three pairs for which the symmetric RBF-SVM confusion increases, as indicated by the black outlines in Fig.~\ref{Fig4}(G). The decline in probe accuracy is thus concentrated in a small number of difficult and visually similar digit pairs. Improvements among the other 42 pairs dominate the average distance-based GDV, whereas classification accuracy is strongly affected by the few local overlaps that generate actual errors.

\NI Taken together, these results show that a decreasing global GDV must not be interpreted as equivalent to monotonically increasing class extractability. The first one or two DBN layers improve both geometric separation and supervised readout, demonstrating a genuine gain in accessible class structure without label-guided representation learning. In higher layers, however, the two properties diverge: most class pairs become more widely separated on average, while information needed to distinguish a few critical pairs is progressively lost. The global GDV faithfully reports the dominant geometric trend, but the probe models and pairwise decomposition reveal whether this trend remains useful for classification. Class separation and class accessibility are therefore complementary rather than interchangeable properties of the internal DBN representations.

\subsection{Origin of enhanced class accessibility}

\NI Looking back at Fig.~\ref{Fig4}(A), the accuracy of all three probes eventually decreases across most of the MNIST DBN layers. Before this decline begins, however, the first hidden layer produces a clear increase in accuracy relative to the original input representation. This early improvement is particularly interesting because H1 contains only 100 neurons instead of 784 input pixels and was trained without class labels. We therefore investigated the H1 effect more closely, extended the analysis to Fashion-MNIST and KMNIST, and perturbed the learned weight structure to determine the origin of the enhanced class accessibility.

\begin{figure}[p]
\centering
\includegraphics[width=\textwidth]{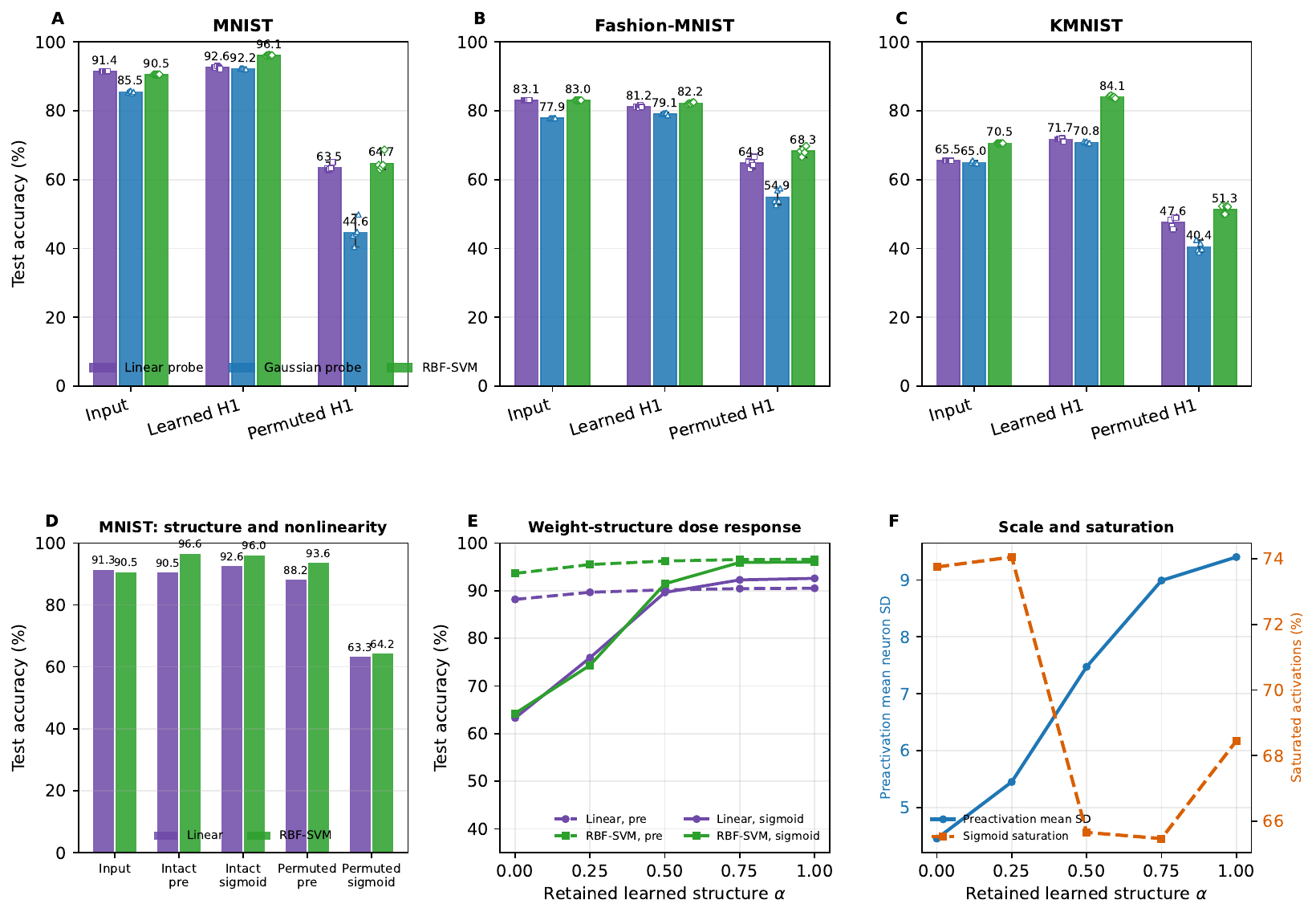}
\caption{{\bf Learned H1 weight structure determines whether nonlinear compression enhances or destroys class accessibility.}
Linear logistic, 200-center Gaussian-kernel, and exact RBF-SVM probes were evaluated on the input, intact H1 activation probabilities, and an H1 control in which every neuron's incoming weights were independently permuted across pixels. \textbf{(A--C)} MNIST, Fashion-MNIST, and KMNIST results across five DBN seeds. Bars show means, open symbols individual seeds, and whiskers minimum--maximum ranges. Permutation preserved each neuron's weight multiset, norm, and bias but caused a pronounced accuracy loss for every dataset and probe. \textbf{(D--F)} Exploratory mechanism analysis for MNIST seed 1234 using linear and RBF-SVM probes. \textbf{(D)} Endpoint comparison of the input and the preactivation (pre) and sigmoid representations generated by intact or permuted weights. Permuted preactivations retain considerable class information, whereas their sigmoid transformation is strongly destructive. \textbf{(E)} Dose response obtained from $W_{\alpha}=\alpha W+(1-\alpha)\widetilde W$, after restoring every column to its learned norm. Accessibility increases systematically with retained learned structure, particularly after the sigmoid. \textbf{(F)} Mean neuronal preactivation standard deviation and the percentage of sigmoid activations below 0.01 or above 0.99. Scale and saturation contribute to, but do not fully explain, the structure-dependent accuracy trend.}
\label{Fig5}
\end{figure}

\NI For each dataset, we compared three representations: the original input pixels, the intact learned H1 activation probabilities, and a structurally perturbed H1 control (Fig.~\ref{Fig5}). The same linear, Gaussian-kernel, and RBF-SVM probes used in Fig.~\ref{Fig4}(A) were fitted independently to every representation. All results were averaged across the five DBN seeds used in the preceding analyses.

\NI For MNIST, the intact H1 representation improves the accuracy of every probe (Fig.~\ref{Fig5}(A)). Linear accuracy increases from approximately 91.4\% on the input pixels to 92.6\% in H1. The Gaussian-kernel probe improves more strongly, from 85.5\% to 92.2\%, and the RBF-SVM rises from 90.6\% to 96.1\%. Thus, the first unsupervised RBM does not merely preserve digit identity during the compression from 784 to 100 variables. It transforms the input into a representation from which digit identity can be recovered more accurately by both linear and nonlinear readouts.

\NI The KMNIST results show an even stronger H1 effect (Fig.~\ref{Fig5}(C)). Linear-probe accuracy increases from approximately 65.5\% to 71.7\%, Gaussian-probe accuracy from 65.0\% to 70.8\%, and RBF-SVM accuracy from 70.5\% to 84.1\%. The largest benefit of the learned H1 representation therefore occurs for the visually more complex Kuzushiji characters. This observation suggests that the first RBM can extract regularities that are difficult to access directly from the raw pixel representation.

\NI The effect is not universal for every dataset and readout. For Fashion-MNIST, the Gaussian probe improves slightly from approximately 77.9\% to 79.1\%, whereas linear accuracy decreases from 83.1\% to 81.2\% and RBF-SVM accuracy from 83.0\% to 82.2\% (Fig.~\ref{Fig5}(B)). The first RBM therefore creates a useful nonlinear reorganization of the Fashion-MNIST representation, but the accompanying compression also discards information used by the linear and RBF-SVM probes. Unsupervised H1 feature learning can enhance class accessibility, but whether this produces a net accuracy gain depends on the structure of the dataset and on the readout.

\NI To test whether the H1 results depend on learned feature structure rather than simply on the marginal distribution or magnitude of the trained weights, we independently permuted the 784 incoming weights of every H1 neuron across pixel coordinates. This operation preserved the exact weight multiset, mean, variance, and norm of each neuron, as well as its hidden bias. It destroyed only the learned assignment of particular weight values to particular image locations. Because the permutation was performed independently for every neuron, spatial stroke-like features and other coordinated pixel patterns were eliminated while the basic per-neuron weight statistics remained unchanged.

\NI Weight permutation causes a pronounced accuracy loss for every dataset and every probe. For MNIST, the linear, Gaussian, and RBF-SVM accuracies fall to approximately 63.5\%, 44.6\%, and 64.7\%, respectively. The corresponding values are 64.8\%, 54.9\%, and 68.3\% for Fashion-MNIST, and 47.6\%, 40.4\%, and 51.3\% for KMNIST. All permuted representations perform substantially worse than both the intact H1 representation and the original input. Moreover, the effect is highly consistent across the five independently trained DBNs, as shown by the individual seed markers in Fig.~\ref{Fig5}.

\NI These perturbation results rule out the possibility that the H1 accessibility is produced merely by the number of hidden neurons, the distribution of trained weight values, their norms, or the hidden biases. Instead, it depends critically on the learned correspondence between image coordinates and neuronal weights. For MNIST and KMNIST, this learned spatial feature structure is the source of the accuracy enhancement over the input representation. For Fashion-MNIST, it still preserves far more class information than a structureless weight arrangement, although the resulting representation does not outperform the input for every probe.

\NI We next separated the learned affine projection from the sigmoid nonlinearity in the seed-1234 MNIST model (Fig.~\ref{Fig5}(D)). For the RBF-SVM, the intact H1 preactivation already increased accuracy from 90.55\% on the input to 96.57\%; applying the sigmoid yielded a similar value of 96.01\%. Hence, the learned linear filters alone make the classes substantially more accessible to this nonlinear readout. For the linear probe, by contrast, the intact preactivation reached 90.50\%, slightly below the input value of 91.33\%, whereas the sigmoid representation reached 92.57\%. The modest linear H1 advantage therefore emerges specifically from applying the nonlinearity to the learned projection.

\NI The corresponding permuted preactivations remained surprisingly informative: they achieved 88.16\% with the linear probe and 93.61\% with the RBF-SVM. The severe loss occurred only after applying the sigmoid, which reduced the accuracies to 63.27\% and 64.23\%, respectively. Thus, permutation does not simply erase digit information at the affine-projection stage. Instead, it changes the projection such that the same sigmoid compression becomes destructive. This finding refines the initial permutation control: the learned spatial weight arrangement is necessary for a useful \emph{nonlinear H1 representation}, but an unstructured 100-dimensional linear projection can still retain considerable class information.

\NI Interpolating continuously between permuted and intact weights produced a systematic dose response (Fig.~\ref{Fig5}(E)). Probe accuracy increased with the retained fraction $\alpha$ of learned weight structure, with the strongest dependence observed after the sigmoid. Because each interpolated weight vector was renormalized to its original norm, this trend cannot be explained by increasing weight magnitude. Preactivation scale and sigmoid saturation nevertheless changed with $\alpha$ (Fig.~\ref{Fig5}(F)). Saturation was highest for fully permuted weights and therefore plausibly contributes to their poor nonlinear representation. However, it was non-monotonic, whereas probe accuracy increased largely monotonically; in particular, performance remained high between $\alpha=0.75$ and $1$ despite increasing saturation. Activation scaling and saturation are consequently contributing factors but do not explain the complete structure-dependent accuracy curve.

\NI Taken together, Fig.~\ref{Fig5} shows that the first RBM can transform raw pixels into compact, class-accessible features without receiving any labels. The learned filters and the sigmoid play distinct but interacting roles: learned affine features already improve nonlinear class accessibility, while the small linear-probe gain over the input requires nonlinear compression of those learned features. The weight structure determines whether that compression is useful or destructive. Particular stroke detectors were not isolated individually, but the interpolation experiment goes beyond their mere necessity and demonstrates a graded dependence of H1 accessibility on the amount of retained learned spatial structure.

\subsection{Layer-wise dreaming and reconstruction ability}

\begin{figure}[p]
\centering
\includegraphics[width=\textwidth,height=0.69\textheight,keepaspectratio]{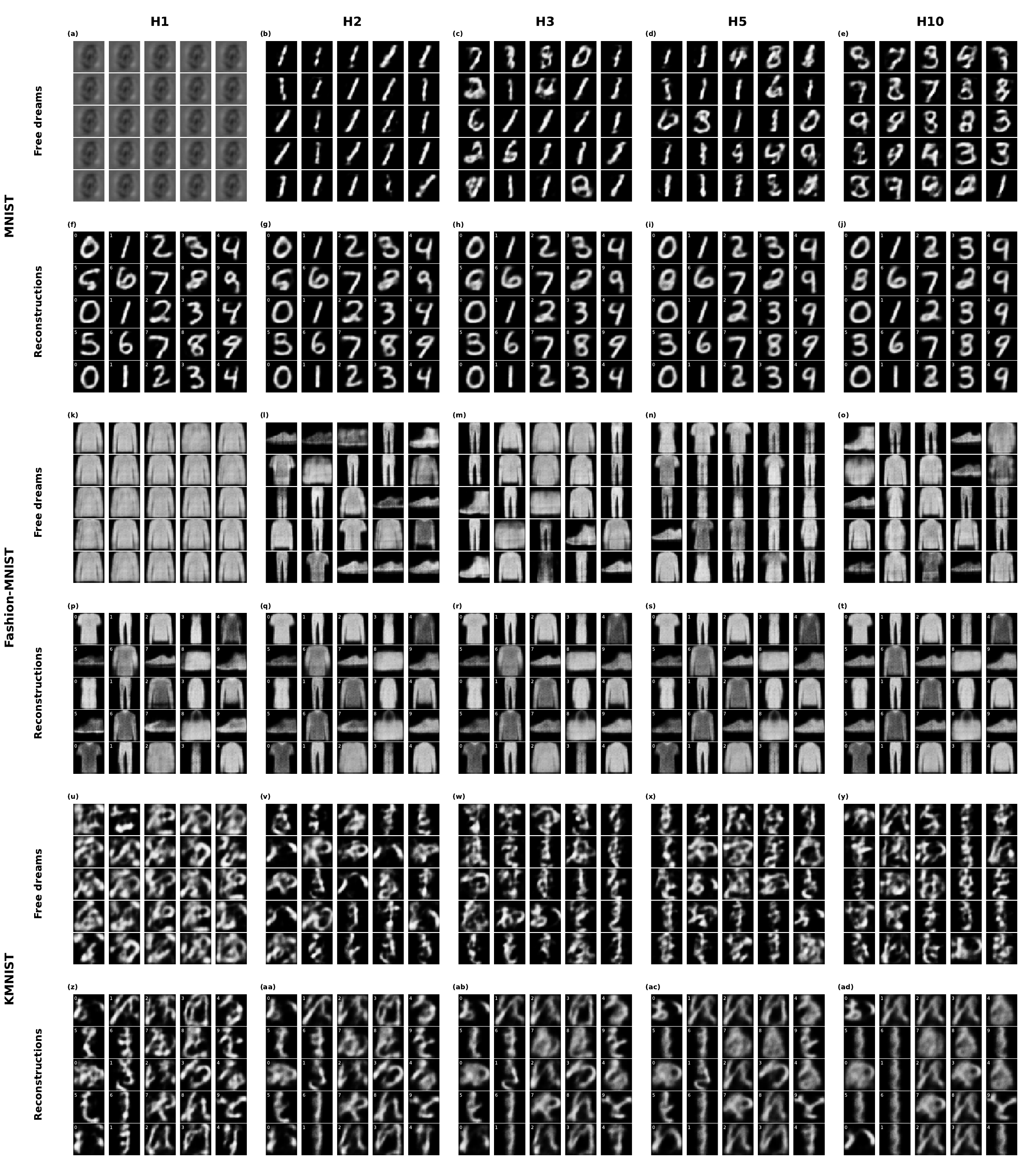}
\caption{{\bf Free generation and data-grounded reconstruction change qualitatively across DBN depth and datasets.}
Ten-layer DBNs with 100 neurons per hidden layer were trained using CD-1 on MNIST, Fashion-MNIST, and KMNIST (seed 1234). Columns show generation or reconstruction through H1, H2, H3, H5, and H10. \textbf{(a--e,k--o,u--y)} For MNIST, Fashion-MNIST, and KMNIST, respectively, 25 independent free-running chains were initialized from Bernoulli$(0.5)$ states in the lower part of the RBM terminating at the indicated layer and evolved for 2,000 alternating Gibbs steps. The resulting states were propagated stochastically through the directed lower DBN; final visible-unit probabilities are shown. No data sample was used to initialize or clamp the chains. \textbf{(f--j,p--t,z--ad)} Twenty-five fixed test images per dataset, with labels cycling through 0--9, were propagated deterministically to the indicated layer and immediately back to the visible layer, without equilibration or sampling. Small numbers denote the true input classes. Free dreams become generally more varied and data-like with depth, but show pronounced dataset-specific mode concentration and ambiguity. Reconstructions retain coarse class identity over much of the hierarchy while progressively losing individual detail and becoming more prototypical.}
\label{Fig6}
\end{figure}

\NI The preceding analyses characterized each hidden layer through aggregate geometric measures and supervised readouts. We finally asked how the same layer-wise transformation appears in image space when the network is used in its generative direction. Figure~\ref{Fig6} contrasts two conceptually different operations: free dreaming, in which no input image constrains the latent state, and data-grounded reconstruction, in which a particular test image is propagated upward to a selected layer and immediately back to the visible units. The first operation tests whether an RBM at a given depth can initiate plausible samples through the lower generative pathway; the second tests how much information about an individual input survives a round trip through that depth.

\NI The data-grounded examples in Fig.~\ref{Fig6}(f--j,p--t,z--ad) provide a visual counterpart to the abstraction distance, operationalized by reconstruction MSE in Fig.~\ref{Fig1}(D--F). Reconstructions through H1 remain close to the supplied images, although thin strokes, textures, and other idiosyncratic details are already smoothed or weakened. With increasing depth, the outputs become less faithful copies and more strongly resemble simplified, class-typical forms. This progression is most immediately visible for MNIST, where many identities remain recognizable through H3 and H5 but H10 occasionally produces ambiguous or apparently changed digits. Fashion-MNIST reconstructions preserve the coarse clothing category comparatively well while progressively homogenizing silhouettes and textures. KMNIST characters undergo stronger blurring and stroke reorganization, consistent with their larger abstraction distance. Across all three datasets, every additional upward and downward transformation discards information that is not recoverable from the selected higher-layer state.

\NI Importantly, the rising abstraction distance does not imply that all class-relevant structure is lost at the same rate as pixel-level detail. As shown by the probe analyses, H1 can improve class accessibility even while its abstraction distance already exceeds zero. The first layer therefore removes some image-specific variation while retaining, and in favorable cases enhancing, features useful for distinguishing classes. At greater depth, however, abstraction distance continues to rise while probe accuracy falls and effective dimensionality collapses. The increasingly prototypical reconstructions are thus not merely denoised versions of the input; they reflect a progressively compressed representation in which both handwriting-specific detail and, eventually, distinctions required for difficult class decisions are lost.

\NI Free dreams show a different and dataset-dependent depth progression (Fig.~\ref{Fig6}(a--e,k--o,u--y)). For MNIST, samples generated through H1 are predominantly unstructured, while H2 is strongly concentrated on a narrow set of one-like forms. H3, H5, and H10 produce a broader range of recognizable digit-like shapes. Fashion-MNIST exhibits an analogous but particularly clear release from mode concentration: H1 generates almost exclusively pullover-like images, whereas H2 and the subsequent depths cover several coarse garment types, including trousers, tops, and shoes. KMNIST dreams are already more heterogeneous at shallow depth and remain visually more ambiguous throughout, but deeper paths increasingly produce structured character-like stroke configurations. Thus, greater depth generally allows a wider repertoire of recognizable samples to emerge, although the trajectory is neither strictly monotonic nor equally successful for all datasets.

\NI These free-dreaming panels are also not stochastic versions of the reconstructions shown below them. Each dream was generated by equilibrating the RBM terminating at the indicated layer without clamping a data example and subsequently propagating its state through the directed lower pathway. The panels consequently probe the generative model associated with different truncation depths rather than a single Gibbs chain equilibrated jointly across all ten layers. Their non-monotonic and dataset-dependent changes---including the shallow mode concentration followed by more varied samples at greater depth---also demonstrate that free-sample quality cannot be inferred directly from abstraction distance, GDV, or probe accuracy.

\NI Taken together, the layer-wise results suggest a coherent but multi-faceted transformation with DBN depth. Early layers trade a modest amount of pixel-level fidelity for a compact feature representation in which class identity can become more accessible. Intermediate layers further suppress input-specific variation and increasingly map images toward class-related, prototype-like structures. In the highest layers, effective dimensionality is strongly reduced, reconstructions lose substantial individual and sometimes class-specific information, and probe accuracy declines, even though the average GDV continues to indicate stronger global class clustering. At the same time, unconstrained latent states can generate more recognizable digit-like patterns through the learned lower hierarchy. Abstraction distance, class accessibility, geometric class separation, dimensional compression, and free generative quality therefore evolve in related but non-equivalent ways; no single one of these quantities provides a complete measure of what the deeper DBN layers have learned.

\section{Discussion}

\subsection{Summary}

\NI The central result of this study is that deep belief networks spontaneously organize their internal representations according to class structure that is never provided during training. Across MNIST, Fashion-MNIST, and KMNIST, the GDV generally decreased with layer depth, showing that examples from the same unknown class became more compact relative to examples from different classes. This behavior was observed across different network widths and was supported by controls excluding untrained transformations, random labels, weight marginals, and sigmoid saturation as sufficient explanations. The progressive class organization therefore reflects statistical structure discovered by the layer-wise generative learning process itself.

\NI The first hidden layers were particularly useful: they frequently made class identity more accessible to both linear and nonlinear probes while reducing 784 input pixels to only 100 latent variables. Thus, the unsupervised RBM did more than preserve information under compression; for MNIST and KMNIST, it reorganized the data into features from which class identity could be extracted more accurately. The learned spatial arrangement of the input weights was essential for this gain. Remarkably, the principal layer-wise curves and probe results also showed very little variation across independently initialized DBNs with different random seeds. The learned representations were therefore not only class-organizing but highly reproducible at the level of the measured population properties, an observation to which we return in the following section.

\NI With increasing depth, the DBNs continued to concentrate the data into compact, class-related structures. Reconstructions became more prototypical, and the effective dimensionality decreased because many responsive neurons acquired similar feature directions and correlated activity patterns. This redundancy can be understood as a consistent convergence onto a comparatively small set of dominant regularities in the data. It also explains how nominally wide layers can implement a much more compact internal code than their neuron count alone would suggest.

\NI Different measures illuminate complementary aspects of this hierarchy. In the higher MNIST layers, the global GDV continued to improve because most class pairs became geometrically better separated, whereas probe accuracy declined because a few difficult pairs lost discriminative detail. Likewise, recognizable free samples emerged along deeper generative paths even though generative quality did not simply track abstraction distance or probe performance. Taken together, the DBN hierarchy performs a progressive unsupervised abstraction: early layers expose class-relevant regularities, and deeper layers consolidate them into increasingly compact and prototype-like representations. This ability to reveal latent class structure without labels is the most striking property of the networks examined here.

\newpage
\subsection{Interpretation}

\NI A broader interpretation of these results is that learning systems can exhibit reproducible macroscopic organization despite enormous freedom in their individual parameters and initial conditions. A previous study of supervised multilayer perceptrons showed that independently initialized networks trained on the same task converge toward task-specific regions of matched weight space \cite{krauss2026convergent}. The present work reveals a complementary phenomenon: without labels or a discriminative objective, independently trained DBNs develop similar layer-wise profiles and progressively recover class structure latent in the input distribution.

\NI In terms of Marr's levels of analysis \cite{marr1982vision}, the image distribution provides computational constraints, the particular weights form an implementation, and the successive feature transformations constitute an intermediate algorithmic organization. Different random seeds produce different microscopic weight realizations, yet GDV, abstraction distance, effective dimensionality, and probe accuracy show remarkably little variation across runs. The input statistics, architecture, and learning rule therefore appear to constrain the algorithmic level much more strongly than the variability of individual parameters would suggest.

\NI Such reproducible solution regions may be described as \emph{algorithmic} or \emph{Platonic attractors}. We use this term descriptively: an attractor denotes a preferred region of parameter or representation space that independently initialized learning systems tend to approach under common constraints, rather than necessarily a single isolated optimum or dynamical fixed point. This view is related to convergent learning \cite{li2015convergent} and to the Platonic Representation Hypothesis, according to which constrained learning systems may discover privileged representations of underlying structure rather than construct arbitrary internal codes \cite{huh2024platonic}.

\NI Two observations are particularly suggestive of such preferred structure. First, the weak seed-to-seed dispersion implies that each DBN depth is associated with a reproducible macroscopic solution. Second, convergence also occurs within individual layers: many neurons in the higher RBMs learn nearly parallel or antiparallel weight vectors and consequently produce strongly correlated responses. Thus, the same feature directions are rediscovered by different neurons within the same system. The progressive dimensional collapse can accordingly be interpreted as many microscopic degrees of freedom becoming organized around a small set of statistically privileged collective features.

\NI Because every RBM is trained on the representation produced below it, the hierarchy may be viewed as a cascade of constrained solution spaces, progressing from relatively local regularities toward fewer and more global feature directions. The unsupervised emergence of class separation is central to this interpretation: digit, garment, and character labels are never supplied, yet their recurring statistical structure is partially recovered by modeling the data distribution itself. At the same time, the differences among GDV, probe accuracy, abstraction distance, and free generation argue against one universal scalar optimum. The attractors are better conceived as structured regions or manifolds expressing characteristic compromises among several constraints. Although the present results do not establish their uniqueness or architecture independence, they show that stochastic learning repeatedly produces collective organization and that the space of viable internal representations is substantially more structured than its vast number of possible neural configurations might suggest.

\subsection{Relation to other work}

\NI The present results build on the original view of restricted Boltzmann machines and DBNs as generative models that can be trained layer by layer with contrastive divergence \cite{hinton2002training,hinton2006fast}. In much of the subsequent literature, however, unsupervised pretraining was chiefly evaluated by how well it initialized a supervised network. Greedy layer-wise training was shown to facilitate the optimization of deep architectures \cite{bengio2007greedy}, and later analyses suggested that it guides learning toward regions of parameter space with favorable generalization properties \cite{erhan2010why}. More broadly, representation learning seeks transformations that expose useful explanatory factors and make downstream tasks easier \cite{bengio2013representation}. Our study adopts a complementary perspective: it leaves the unsupervised hierarchy frozen and directly follows how class geometry, class accessibility, abstraction distance, dimensionality, and free generation evolve from one layer to the next.

\NI The reduction of representational dimensionality is related to earlier work on deep autoencoders, which demonstrated that nonlinear multilayer networks can learn compact codes that preserve important structure in high-dimensional data \cite{hinton2006reducing}. Low intrinsic dimensionality has also been observed in supervised deep networks, where the dimensionality of hidden representations can be far smaller than the number of units and often decreases toward the output \cite{ansuini2019intrinsic}. The present findings extend this picture to greedily trained, fully unsupervised DBNs. Moreover, the neuron-level analyses identify a concrete origin of the decrease: higher layers do not simply contain increasing numbers of inactive units; rather, many responsive neurons acquire aligned or anti-aligned weight vectors and correlated activation profiles, so that a nominally wide layer becomes dominated by comparatively few collective feature directions.

\NI The increasingly compact class geometry may appear reminiscent of \emph{neural collapse}, the terminal-phase organization found in supervised classifiers, in which within-class variability contracts and class means approach a highly symmetric configuration \cite{papyan2020prevalence}. The analogy is informative but limited. Neural collapse is driven by labeled discriminative training near the output of a classifier, whereas the present effect develops throughout a hierarchy trained only to model its input distribution. We also do not observe or claim the specific simplex geometry associated with neural collapse. Most importantly, GDV and probe accuracy can move in different directions in the higher layers. The DBN phenomenon is therefore better described as unsupervised class-related organization and compression than as neural collapse in its original sense.

\NI Our weak seed-to-seed variation and the repeated learning of similar features also connect to work on convergent representations. Independently trained neural networks can recover corresponding features or representational subspaces \cite{li2015convergent}, and centered kernel alignment can reveal reproducible layer correspondences across different initializations \cite{kornblith2019similarity}. The broader Platonic Representation Hypothesis proposes that sufficiently constrained learning systems may converge toward common representations of underlying reality \cite{huh2024platonic}. The present results add two observations at a smaller and more mechanistic scale: independently initialized DBNs follow highly reproducible layer-wise trajectories, and multiple neurons within a single unsupervised layer repeatedly discover the same dominant feature directions.

\NI At the same time, the emergence of label-related structure should not be interpreted as unrestricted recovery of uniquely defined semantic factors. In the general case, unsupervised disentanglement is not identifiable without inductive biases concerning the model or the data \cite{locatello2019challenging}. Here, the image distribution, RBM architecture, binary stochastic units, contrastive-divergence rule, and layer-wise training procedure provide precisely such constraints; class labels enter only after training as an external diagnostic. Our conclusion is consequently narrower but directly supported by the experiments: under these conditions, generative learning reproducibly reorganizes several image datasets in a direction that exposes part of their latent class structure.

\NI Taken together, previous studies have established the generative capability of DBNs, the optimization benefits of unsupervised pretraining, dimensional compression in deep representations, and convergence among independently trained networks. The distinctive contribution of the present work is to examine these themes jointly along an entirely unsupervised DBN hierarchy. This comparison reveals that geometric class separation, probe accessibility, abstraction distance, effective dimensionality, and free-sample quality are related but non-equivalent properties. Their layer-wise dissociation provides a more differentiated account of what unsupervised depth contributes than any one reconstruction or classification score alone.

\subsection{Outlook}

\NI An immediate question is whether the decline of probe accuracy in the higher DBN layers reflects an actual loss of class information or only a loss of information that is accessible to the probe families considered here. This distinction is particularly relevant for H10: its two-dimensional MDS projection still exhibits substantial class-related organization, although all three probes perform worse than in the earlier layers. MDS cannot resolve this issue because it is a low-dimensional visualization that may emphasize broad cluster structure while obscuring difficult local boundaries. A natural next step is therefore to fit regularized multilayer perceptrons of systematically increasing capacity to every frozen layer. Equal training sets, nested hyperparameter selection, learning curves, and an untouched common test set would be essential. If increasingly expressive probes consistently recovered more test-set information from H10 than from H1 or even the input, this would demonstrate that deep DBN representations contain useful but nonlinearly encoded class structure. Conversely, high training accuracy without a corresponding test advantage would indicate memorization rather than generally accessible information, while saturation below the early-layer performance would strengthen the case for genuine information loss.

\NI A second direction follows directly from the redundancy observed in Fig.~\ref{Fig3}. If several neurons within one layer learn nearly the same feature, the nominal layer width considerably overstates the number of distinct features required by the data. This suggests replacing fixed-width DBNs by an adaptive procedure that alternates learning with consolidation. Neurons could first be grouped using both weight-vector similarity and activation correlation; redundant members could then be merged or removed, followed by a short unsupervised retraining phase before the next RBM is added. Weight similarity alone would not be sufficient, because opposite weight directions, different biases, or similar filters operating in different activation regimes need not be functionally interchangeable. The decisive criterion should instead be whether removal preserves the modeled distribution, reconstruction behavior, and representational geometry on held-out data.

\NI An even more attractive variant would determine layer width online. Inspired by the match-based recruitment principle of Adaptive Resonance Theory \cite{carpenter1987massively}, a new neuron or feature channel could be introduced only when the existing feature set fails to represent a sufficiently novel input pattern, whereas repeated solutions would be consolidated rather than stored as additional copies. This would not be ART in the strict sense, since the units would remain features of an RBM rather than explicit recognition categories. Nevertheless, an analogous novelty or vigilance criterion could turn layer width from a manually chosen hyperparameter into an emergent property of the input distribution and learning stage. Comparing such networks with conventionally trained DBNs across a dense range of fixed widths would reveal whether each hierarchy level possesses a reproducible minimal feature basis and whether adaptive compression can retain the favorable GDV trajectory without unnecessary multiplicity.

\NI The exploratory variants considered during development of this study also suggest that preventing redundancy requires more than keeping individual neurons active. Global lateral inhibition, random sparse connectivity, restricted topography, and a common target activity did not by themselves guarantee diverse feature learning, whereas weight normalization provided only partial stabilization. Future mechanisms should therefore act directly on functional overlap, for example through local competition among similarly responding neurons, anti-Hebbian lateral plasticity, diversity penalties, or novelty-triggered recruitment and pruning. Such interventions should remain label-free during training and be evaluated jointly by effective dimensionality, weight and activity redundancy, abstraction distance, reconstruction characteristics, GDV, and probe performance. The aim would not be to maximize dimensionality, but to preserve the smallest diverse representation that retains the useful structure of the data.

\NI Finally, the relation between local RBM learning and the global generative model remains open. Bottom-anchored negative phases, in which locally generated states are decoded through the lower stack and encoded again, could test whether enforcing lower-stack consistency changes the balance between class organization and reconstruction. Joint fine-tuning or a fully coupled deep Boltzmann formulation would provide stronger but less local alternatives. These models should be assessed not only by pixel-wise abstraction distance (reconstruction MSE), but also by approximate likelihood or free-energy-based criteria and by quantitative measures of the fidelity, class coverage, and diversity of freely generated samples. Extending the present analysis to more heterogeneous natural data would then show whether unsupervised class organization, feature convergence, and adaptive dimensional compression are general properties of hierarchical generative learning or are favored by the comparatively regular image datasets studied here.

\section{Additional Information}

\subsection{Author contributions}

\NI CM conceived the study, implemented the methods, evaluated the data, and wrote the paper. 
AS discussed the results and acquired funding.
AM and TK discussed the results and provided resources.
PK conceived the study, discussed the results, acquired funding and wrote the paper.

\subsection{Funding}
\NI This work was funded by the Deutsche Forschungsgemeinschaft (DFG, German Research Foundation): grants KR\,5148/3-1 (project number 510395418), KR\,5148/5-1 (project number 542747151), KR\,5148/10-1 (project number 563909707) and GRK\,2839 (project number 468527017) to PK, and grants SCHI\,1482/3-1 (project number 451810794) and SCHI\,1482/6-1 (project number 563909707) to AS.

\subsection{Competing interests statement}
\NI The authors declare no competing interests.

\subsection{Data availability statement} 
\NI The complete data and analysis programs will be made available upon reasonable request.

\subsection{Third party rights}
\NI All material used in the paper are the intellectual property of the authors.


\newpage
\bibliographystyle{unsrt}
\bibliography{references}


\clearpage

\clearpage

\clearpage

\clearpage

\clearpage

\clearpage

\clearpage

\end{document}